\documentclass{elsart}
\journal{New Astronomy}

\usepackage{epsfig}

\def\url#1{{\ttfamily\def\/{/\discretionary{}{}{}}#1}}

\begin{document}

\begin{frontmatter}
\title{Separation of foreground and background signals in single frequency measurements
of the CMB Polarization}
\author[moscow]{M.V.Sazhin\thanksref{fn1}},
\author[mibic]{G.Sironi\thanksref{fn2}},
\author[moscow]{O.S.Khovanskaya\thanksref{fn3}}
\address[moscow]{Sternberg Astronomical Institute, Universitetsky pr. 13,
119899 Moscow Russia}
\address[mibic]{Dipartimento di Fisica G. Occhialini - University
of Milano Bicocca - Piazza della Scienza 3 - I20126 Milano -Italy}
\thanks[fn1]{E-mail: sazhin@sai.msu.ru}
\thanks[fn2]{E-mail: giorgio.sironi@mib.infn.it}
\thanks[fn3]{E-mail: khovansk@sai.msu.ru}

\begin{abstract}
The polarization of the Cosmic Microwave Background (CMB)is a powerful
observational tool at hand for modern cosmology. It allows to break the
degeneracy of fundamental cosmological parameters one cannot obtain using
only anisotropy data and provides new insight into conditions existing in the
very early Universe. Many experiments are now in progress whose aim is
detecting anisotropy and polarization of the CMB. Measurements of the CMB
polarization are however hampered by the presence of polarized foregrounds,
above all the synchrotron emission of our Galaxy, whose importance increases
as frequency decreases and dominates the polarized diffuse radiation at
frequencies below $\simeq 50$ GHz. In the past the separation of CMB and
synchrotron was made combining observations of the same area of sky at
different frequencies. In this paper we show that the statistical properties
of the polarized components of the synchrotron and dust foregrounds are
different from the statistical properties of the polarized component of the
CMB, therefore one can build a statistical estimator which allows to extract
the polarized component of the CMB from single frequency data also when the
polarized CMB signal is just a fraction of the total polarized signal.  Our
estimator improves the signal/noise ratio for the polarized component of the
CMB and reduces from $\simeq$50 GHz to $\simeq$20 GHz the frequency above
which the polarized component of the CMB can be extracted from single
frequency maps of the diffuse radiation.
\end{abstract}

\begin{keyword}
Polarimetry, Mathematical Procedures, Radio and microwave, Observational
Cosmology
\PACS: 95.75.Hi, 95.75.Pq, 95.85.Bh, 90.80.Es
\end{keyword}
\end{frontmatter}

\section{Introduction}

Almost a decade elapsed since the first detection of the anisotropy of the
Cosmic Microwave Background at large angular scales ($\ge 10^0$) \cite{str92},
\cite{smo92}. Today the CMB anisotropy (CMBA) has been detected also at
intermediate ($\sim (1^0 - 10^0)$) and small angular scales ($\le 1^0$), so
the CMBA angular spectrum is now reasonably known  down  to  the  region of
the ~first  ~and   ~second   ~Doppler ~peaks ~\cite{boo00}, ~\cite{boo01a},
\cite{boo01b}. Its shape gives information e.g. on the spectrum of the
primordial cosmological perturbations or can be used to test the inflation
theory but rises new questions to which CMBA cannot answers. It is however
possible to get answers looking ~at ~the ~CMB ~polarization ~(CMBP). For
instance one can use CMBP to disentangle the effects of fundamental
cosmological parameters like density of matter, density of dark energy etc.
This is among the goals of space and ground based experiments like \cite{Map},
\cite{Planck1},  ~\cite{Planck2},  ~\cite{ami},  ~\cite{que},  \cite{gerva},
\cite{newboo} and is the main goal of SPOrt a polarization dedicated ASI/ESA
space mission on the International Space Station \cite{cor99}.

 The relevance
of the CMB polarization was remarked for the first time by M. Rees
\cite{ree65}. Since his paper many models of the expected features of the
CMBP have been published (see for instance \cite{saz95}, \cite{ng96},
\cite{mel97}). ~They ~immediately ~stimulated ~the ~search ~for CMBP, but the
first detection has been obtained only a few months ago (\cite{dasi} and
\cite{map1}). Because of its importance this discovery must be confirmed by
new observations made with different systems and using different methods of
extraction of the CMBP from the sky signal. The detection of the CMBP is in
fact extremely difficult because the signal is at least an order of magnitude
below the CMBA level. Moreover polarized foregrounds of galactic origin may
cover the CMBP and/or mimic CMBP spots by their inhomogeneities. The signal
to noise (CMBP $/$ polarized foreground) ratio is therefore $<< 1$. In this
paper we discuss a method which improves this ratio and allows to disentangle
CMBP and polarized foregrounds.

In the microwave range the galactic foregrounds include:
\begin{itemize}
\item synchrotron radiation (strong polarization),
\item free-free emission (null or negligible polarization),
\item dust radiation (polarization possible)
\end{itemize}

Beccause here we are interested in polarization, in the following  we will
neglect free-free emission.The effects of dust, if present, (e.g.
\cite{set98}, \cite{fos01}), will be added to the synchrotron effects. In
fact, as it will appears in the following, what matters in our analysis are
the statistical properties of the spatial distribution of the foregrounds and,
by good fortune, the spatial distribution of the dust polarized emission is
similar to that of the synchrotron emission. Behind both types of radiation
there is in fact the same driving force, the galactic magnetic field which
alignes dust grains and guides the radiating electrons (\cite{Cor03} and
references therein).

For the anisotropy the separation of foregrounds and CMB ~was ~successfully
solved  ~by  ~the DMR/COBE  ~team  ~when  ~they  ~discovered  ~~CMBA
\cite{smo92}. The separation of CMBP and foregrounds is more ~demanding ~and
recognizing true CMB spots among foreground inhomogneities severe.
~Approaches ~~commonly ~~used ( e.g. ~\cite{dod97}, ~~\cite{teg99},
~~~~\cite{sto01},  ~~~\cite{teg96}, ~~~\cite{kog00}, \\ ~\cite{map2}), are
based on the differences between the frequency spectra of foregrounds and CMB,
therefore require multifrequency observations.

In this paper we suggest a different method. It takes advantage of the fact
that, as we will show in the following, at small angular scales the values of
the parameters used to describes the polarization of the diffuse radiation at
a given frequency in different directions fluctuate.  We propose of analyzing
the angular distribution of the polarized radiation on single frequency maps
of the diffuse radiation and disentangling the main components, polarized
synchrotron and CMBP, looking to their different statistical properties. This
method was proposed and briefly discussed in \cite{saz01}. Here we present a
more complete analysis.

\section{Polarization Parameters}
\subsection{The Stokes Parameters}
Convenient ~quantities ~~commonly ~~used ~~to ~~describe ~the ~~polarization
~status ~~of ~~radiation ~~are the Stokes parameters ~~(see ~for ~instance \\
~\cite{gin64}, ~\cite{gar66}, \\ ~\cite{gin69}).

Let's assume a monochromatic plane wave of intensity I and amplitude $\propto
\sqrt {I}$. In the {\it observer plane}, orthogonal to the direction of
propagation of the electromagnetic wave, we can choose a pair of orthogonal
axes ${\vec l}$ and ${\vec r}$. On that plane the amplitude vector of an
unpolarized wave moves in a random way. On the contrary it describes a
figure, the {\it polarization ellipse}, when the wave is polarized.

Projecting the wave amplitude on ${\vec l}$ and ${\vec r}$ we get two
orthogonal, linearly polarized, waves of intensity $I_l$ and $I_r$,($I = I_l
+ I_r$) whose amplitudes are $\propto \sqrt {I_l}$ and $\propto \sqrt {I_r}$
respectively. If the original wave of intensity $I$ is polarized, $I_l$ and
$I_r$ are correlated: let`s call $I_{12}$ and $I_{21}$ their correlation
products.

By definition the Stokes parameters are: $I=I_l +I_r$, $Q=I_l -I_r$,
$U=I_{12} + I_{21}$, and $V= i(I_{21} - I_{12})$.

$Q$ and $U$ describe the linear polarization, $V$ the circular polarization
and $I$ the total intensity.

\noindent The ratio

\begin{equation}
\tan 2 \chi = \frac{U}{Q}
\end{equation}

\noindent gives the angle $\chi$ between the vector $ \vec l$ and the main
axis of the polarization ellipse ($0 < \chi < \pi$).

Rotation by an angle $\phi$ of the $(l,r)$ coordinate system give a new
coordinate system $(\hat l , \hat r)$ in which the Stokes parameters become

\begin{equation}
\begin{array}{c}
\hat Q = Q \cos 2\phi -U \sin 2\phi \\
~\\ \hat U = U \cos 2\phi + Q \sin 2\phi
\end{array}
\end{equation}

\begin{figure}
\vspace{-15pt} \centerline{\hspace{30pt}\epsfig{file=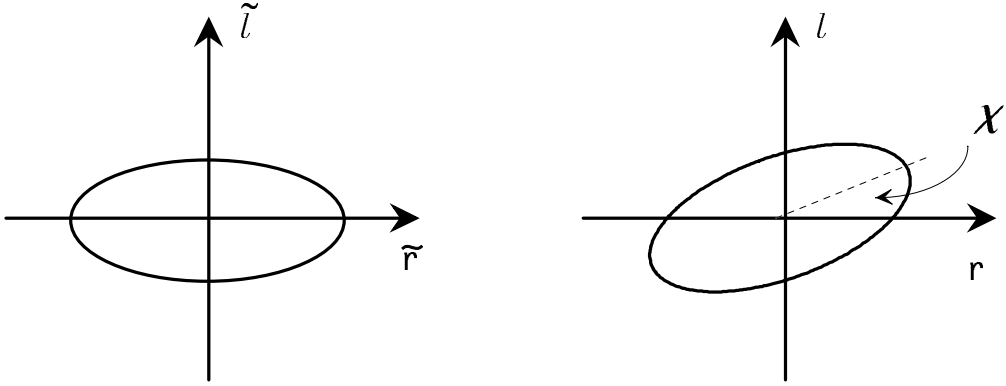,height=2in}}
\vspace{-12pt} \caption{\small The polarization ellipse} \label{f2}
\end{figure}

\noindent So when $\phi = \chi =(1/2) \arctan {\frac{U}{Q}} $ the axes of the
polarization ellipse coincide with the reference axes $\hat l$ and $\hat r$
(see fig.\ref{f2}).

\subsection{Electric and Magnetic Modes}

To analyze the properties of the CMB polarization it is sometimes convenient
to use rotationally invariant quantities, like the radiation intensity $I$ and
two combinations of $U$ and $Q$: $Q+iU$ and $Q-iU$. The intensity $I$ can be
decomposed into usual (scalar) spherical harmonics $Y_{lm}(\theta, \varphi)$.
\begin{equation}
I=\sum_{l,m} a_{lm} Y_{lm}(\theta, \varphi)
\end{equation}

\noindent The quantities $Q \pm iU$ ~can ~be ~~decomposed ~into ~~$\pm 2$
~spin harmonics \\ \cite{saz96a}, ~~\cite{saz96b}, ~~\cite{sel97} \\ $
Y_{lm}^{\pm 2}(\theta, \varphi)$ \footnote{Alternatively, one can use the
equivalent polynomials derived in \cite{gol67}} :
\begin{equation}
Q \pm iU=\sum_{l,m} a_{lm}^{\pm 2} Y_{lm}^{\pm 2}(\theta, \varphi)
\end{equation}

The  $\pm 2$ spin harmonics form a complete orthonormal system (see, for
instance,  \cite{gol67},\cite{gel58}, \cite{zer70}, \cite{tho80}) and can be
written \cite{saz96b}, \cite{saz99}:
\begin{equation}
\begin{array}{c}
Y^2_{lm}(\theta, \varphi) = N^2_{lm} P^2_{lm}(\theta) e^{i m \varphi} \\
~\\ Y^{-2}_{lm}(\theta, \varphi) = N^{-2}_{lm} P^{-2}_{lm}(\theta) e^{i m
\varphi}
\end{array}
\end{equation}

\noindent where
\begin{equation}
P^s_{lm}(x) = (1 - x)^{\displaystyle (m + s) \over 2} (1 + x)^{\displaystyle
(s - m) \over 2} P^{(m + s, s - m)}_{l-s}(x) \label{jac}
\end{equation}
\noindent is a generalized Jacobi polynomial, $s = \pm 2$ and:
$$
N^s_{lm} = \frac{1}{2^s}\sqrt{\frac{2l + 1}{4 \pi}} \sqrt{\frac{(l - s)!(l +
s)!}{(l - m)! (l + m)!}}
$$
\noindent is a normalization factor.

The harmonics amplitudes $a^{\pm 2}_{lm}$ correspond to the Fourier spectrum
of the angular decomposition of rotationally  invariant combinations of Stokes
parameters.

Because spin $\pm 2$ spherical functions form a complete orthonormal system:
\begin{equation}
\int_{4 \pi} Y^{\pm 2}_{l m}(\theta, \varphi)Y^{*\; \pm 2}_{l^\prime
m^\prime}(\theta, \varphi) d\Omega = \delta_{l m}\delta_{l^\prime m^\prime}
\label{orth}
\end{equation}

\noindent  we can write
\begin{equation}
a^{\pm 2}_{l m} =  \int_{4 \pi} d\Omega \; \; \left(Q(\theta, \varphi) \pm i
U(\theta, \varphi)\right) Y^{*\; \pm 2}_{l m} \label{ampl}
\end{equation}

Following \cite{sel97} and the very nice introduction made more recently by
\cite{zal01} we now introduce the so called {\it $E$ (electric)} and {\it $B$
(magnetic) modes} of these harmonic quantities:
\begin{equation} \begin{array}{c}
a^E_{lm} = \frac{1}{2}\left( a^{+2}_{lm} + a^{-2}_{lm}\right) \\
~\\ a^B_{lm}=\frac{i}{2}\left( a^{+2}_{lm} - a^{-2}_{lm}\right)
\end{array} \end{equation}

They have different parities. In fact when we transform the coordinate system
$Oxyz$ into a new coordinate system $\tilde O \tilde x \tilde y \tilde z$,
such that

\begin{equation}
\begin{array}{c}
\tilde  {\vec l} = \vec l \\
\tilde  {\vec r} = - \vec r
\end{array}
\end{equation}

\noindent the E and B modes transform in a similar way:
\begin{equation}
\begin{array}{c}
\tilde a^{E} = a^{E} \\
\tilde a^{B} =- a^{B}
\end{array}
\end{equation}

\noindent $Q$ remains identical in both reference systems and $U$ changes
sign.

It is important to remark that $a^E$ and $a^B$ are uncorrelated.

In terms of $Q$ and $U$ we can write:
\begin{equation} \begin{array}{c}
a^E_{l m} =\frac{1}{2} \int d\Omega \; \; \left(Q \left( Y^{+2}_{l m} +
Y^{-2}_{l m}\right) + iU\left( Y^{+2}_{l m} - Y^{-2}_{l m} \right) \right) \\
~\\ a^B_{l m} =\frac{1}{2} \int d\Omega \; \; \left(iQ \left( Y^{+2}_{l m} -
Y^{-2}_{l m}\right) - U\left( Y^{+2}_{l m} + Y^{-2}_{l m} \right) \right)
\end{array} \end{equation}

\noindent therefore :
\begin{equation}
\begin{array}{c}
\left<(a^E_{l m})^2\right> - \left<(a^B_{l m})^2\right> = \\
~~~~~=~2 \int d\Omega \left(\left<|Q^2|\right> - \left<|U^2|\right>\right)
\left(Y^{+2}_{l m} Y^{* -2}_{l m} + Y^{* +2}_{l m} Y^{ -2}_{l m}\right)
\label{syn1}
\end{array}
\end{equation}
Here $\left<|Q^2|\right>$ and $\left<|U^2|\right>$ designate values of delta
correlated 2D stochastic fields $Q$ and $U$. Omitting mathematical details,
the correlation equations  for $Q$ and $U$ are:
\begin{equation}
\begin{array}{c}
\left<Q Q^*\right> = |Q^2| \delta(\Omega - \Omega^\prime) \\
~\\ \left<UU^*\right> = |U^2| \delta(\Omega - \Omega^\prime) \\
~\\ \left<QU^*\right>= 0
\end{array}
\end{equation}
\noindent where $\delta(\Omega - \Omega^\prime) =\delta(\cos \theta -\cos
\theta^\prime) \cdot \delta(\varphi - \varphi^\prime)$ is the Dirac delta -
function on the sphere. (In the following we will sometimes omit indexes $l$
and $m$).

\section{Synchrotron Radiation and its Polarization}

Synchrotron radiation results from the helical motion of extremely
relativistic electrons around the field lines of the galactic magnetic field
(see, for instance \cite{gin64}, \cite{gin69}, \cite{wes59}). The electron
angular velocity

\begin{equation}
\omega_e =\frac{e H_p}{m_ec}~\frac{m_e c^2}{\mathcal{E}} = \omega_o \frac{m_e
c^2}{\mathcal{E}} \label{gyr1}
\end{equation}

\noindent is determined by the ratio between $H_p$, the component of the
magnetic field orthogonal to the particle velocity, and $\mathcal{E}$, the
electron energy. As it moves around the magnetic field lines the electron
radiates.
\par~\par
\noindent a){\it Single electron}
\par
\begin{figure}
\vspace{-15pt} \centerline{\hspace{30pt}\epsfig{file=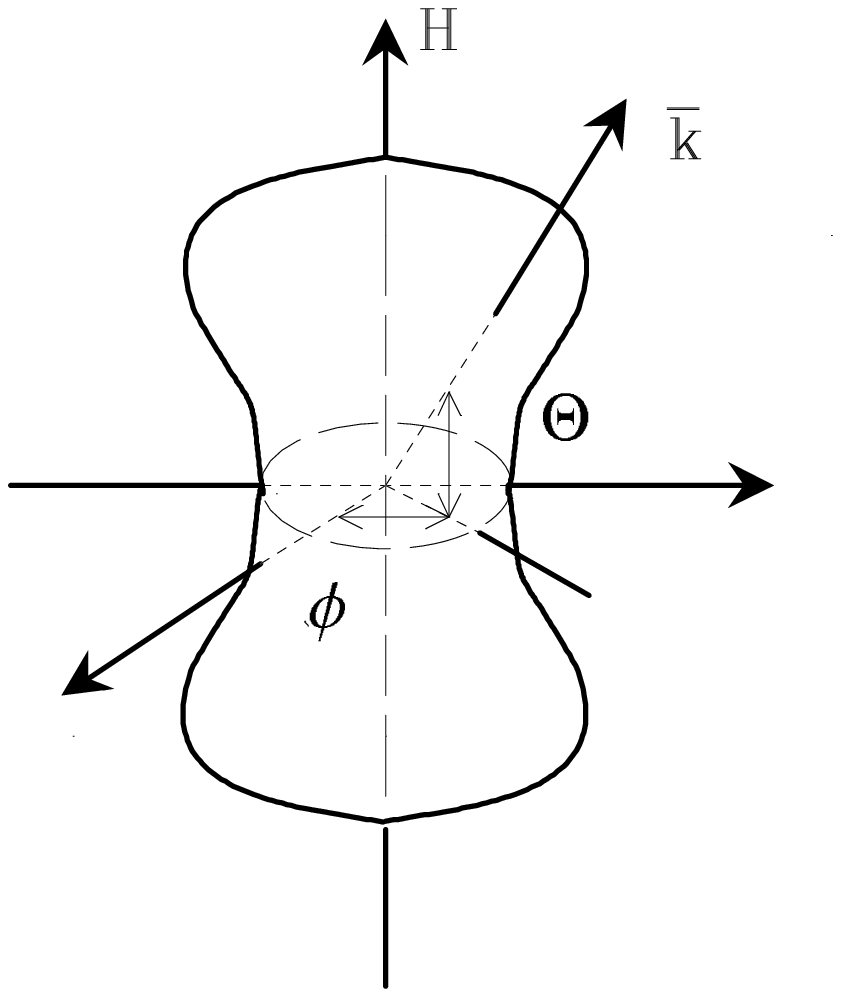,height=2in}}
\vspace{-12pt} \caption{\small Spatial distribution of the cyclotron
radiation produced by a single electron (see eq.\ref{dumbb}) } \label{f3}
\end{figure}

\noindent Until the electron velocity is small ($v<<c$) we speak of cyclotron
radiation: the electron behaves as a rigid dipole which rotates with
gyrofrequency (\ref{gyr1}) in a plane orthogonal to the magnetic field
direction and emits a single line. The radiation has a dumbell spatial
distribution (see fig.\ref{f3}):
\begin{equation}
I(\Theta, \Phi) \sim (1 + \cos^2 \Theta), \label{dumbb}
\end{equation}
\noindent is circularly polarized along the dumbell axis ($\Theta = 0$) and
linearly polarized in directions orthogonal to it ($\Theta = 90^{\circ}$).

\par When the electron velocity increases the radiation field changes
until at $v \approx c$, ($\mathcal{E}$ $\gg m_e c^2$) it assumes the peculiar
charachters of synchrotron radiation:

i)radiated power proportional to $\mathcal{E}$ $^2$ and $H_p^2$,

ii)continous frequency spectrum, peaked around:
\begin{equation}
\omega_m(\psi) = \omega_o \frac{\sqrt{1 - \frac{v^2}{c^2}}}{1 -
\frac{v}{c}\cos\psi} ~\propto H_p{\mathcal{E}} ^2
\end{equation}
\par
\begin{figure}
\vspace{-15pt} \centerline{\hspace{30pt}\epsfig{file=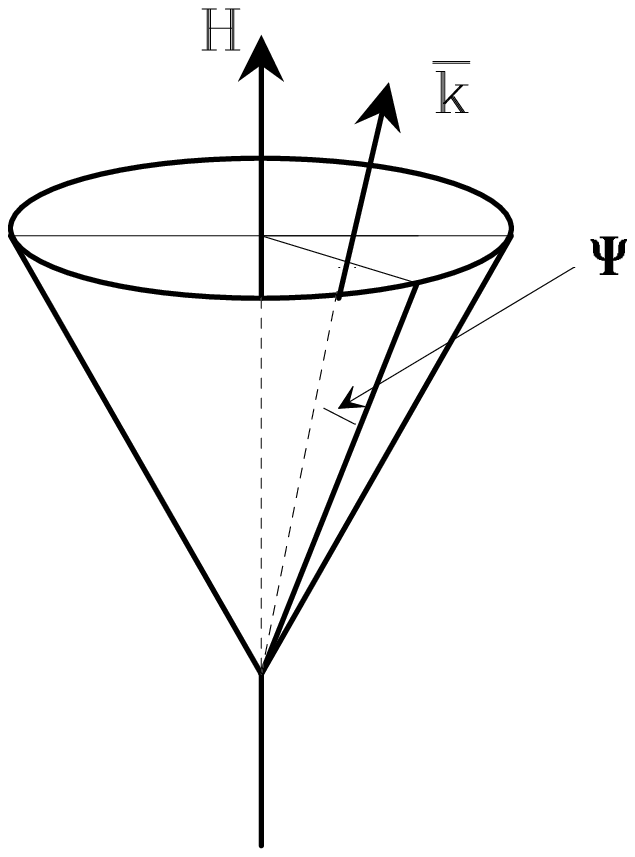,height=2in}}
\vspace{-12pt} \caption{\small Velocity cone of an ultrarelativistic
($v\simeq c$) electron  ($H$ = magnetic vector, $k$ wave vector, $\psi$ angle
between the electron velocity and the direction of observation)} \label{f4}
\end{figure}

\begin{figure}
\vspace{-15pt} \centerline{\hspace{30pt}\epsfig{file=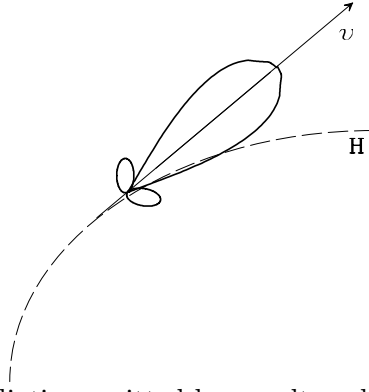,height=2in}}
\vspace{-12pt} \caption{\small Distribution of the radiation emitted by an
ultrarelativistic ($v \simeq c$) electron with istantaneous velocity $v$
spiralling (dashed line) around the lines of the magnetic field $H$
(orthogonal to the sheet)} \label{f5}
\end{figure}

\noindent ($\psi$ is
the angle between the velocity vector $\vec v$ and the wave vector $\vec k$,
see fig.\ref{f4}). The peak is so narrow  that in a given direction the
emission is practically monochromatic,

iii)radiation almost entirely emitted in a narrow cone   \footnote{the
symmetry plane($\Theta = 90^{\circ}$) of the cyclotron dumbbell beam, seen by
a fast moving observer ($v \approx c$) becomes a cone folded around the
direction of movement} of aperture (see fig.\ref{f5})
\begin{equation}
\psi \approx \frac{m_e c^2}{\mathcal{E}}
\end{equation}
\par\noindent around the forward direction
of the electron motion. Inside the cone ($\cos \psi \approx 1$) the frequency
is maximum and equal to
\begin{equation}
\omega_{m,o} \approx \omega_o \left(\frac{{\mathcal{E}}}{m_e c^2}\right)^2,
\end{equation}
\noindent In the opposite direction ($\cos \psi \approx -1$) intensity and
frequency are sharply reduced.

iv)radiation 100 $\%$ linearly polarized at the surface of the cone.Inside
the cone the linear polarization is still dominant but a small fraction of
circular polarization exists, ($V \sim o(\frac{m_e c^2}{E})$ and $I \sim U
\sim Q$). Outside the cone the very small fraction of radiation produced is
elliptically polarized and becomes circularly polarized when seen along the
$H_p$ direction. So the Stokes parameters depend on $H$, the angle $\mu$
between $H$ and the line of sight, and the dimensionless frequency
$\nu/\nu_c$, where $\nu_c = 1.5(\omega_{m}(0)/2\pi)$ is the so called {\it
critical frequency}, \cite{gin64},\cite{gin69}, itself function of $H$ (see
eq.\ref{gyr1}).

\par~\par
\noindent b){\it Cloud of monoenergetic electrons}
\par\noindent When the effects of many monoenergetic electrons with
uniform distribution of pitch angles are combined, $I, Q$ and $U$ are
reinforced (the Stokes parameters are additive) while $V$ is erased. In fact

\begin{equation}
\begin{array}{c}
I(\nu) =c_1 H_p F(\frac{\nu}{\nu_c}) \\
~\\ Q(\nu)=c_2H_pF_p(\frac{\nu}{\nu_c}) \cos 2\chi  \\
~\\ U(\nu)=c_2H_pF_p(\frac{\nu}{\nu_c}) \sin 2\chi  \\
~\\ V(\nu) \approx 0
\end{array}
\end{equation}

\noindent where $c_1$ and $c_2$ are constants, $F$ and $F_p$ almost
monochromatic functions of $\nu$ , and $\chi$ the angle between the projection
of the magnetic vector on the observer plane and an axis on that plane (The
projection of the magnetic vector on the observer plane is the minor axis of
the polarization ellipse). If inside the emitting cloud the magnetic field
varies, also $\chi$ varies, therefore $\cos 2\chi$ and $\sin 2\chi$ must be
averaged along the line of sight across the cloud. In conclusion the degree of
polarization $p$ varies between a maximum value, (uniform magnetic field) and
zero (magnetic field randomly distributed).
\par~\par
\noindent c){\it Electrons with power law energy spectrum}

In the interstellar medium the radiating particles are the cosmic ray
electrons whose energy spectrum :
\begin{equation}
N(\mathcal{E}) = K \mathcal{E}^{- \gamma}
\end{equation}

\noindent is a power law (see for instance \cite{ele1}, \cite{ele2}) and
references therein) with spectral index  $\gamma \approx 2.4 - 3.0$ and space
density proportional to $K$. Because the emission of a single electron is
practically monochromatic the resulting radiation spectrum is a power law:
\begin{equation}
I(\nu) = I_0(\gamma) H_p^{\frac{\gamma +1}{2}} \nu^{-\beta'} \label{spec1}
\end{equation}
\noindent with intensity spectral index $\beta' = \frac{\gamma-1}{2}$,
temperature spectral index  $\beta = \beta' + 2$ and $I_0$ a slow function of
$\gamma$ \cite{gin64}. (If the magnetic field is not uniform  we use $<H_p>$
instead of $H_p$ and  a slightly different function $I_0(\gamma)$).

\noindent The Stokes parameters are products of the intensity $I(\nu)$, the
degree of polarization $p$ and $\cos 2\chi$ or $\sin 2\chi$, therefore we can
write:
\begin{equation}
\begin{array}{c}
Q= c_3~H_p^{\frac{\gamma + 1}{2}} ~\nu^{-\frac{\gamma -1}{2}} \cos 2 \chi \\
~\\U=c_3~H_p^{\frac{\gamma + 1}{2}} ~\nu^{-\frac{\gamma-1}{2}} \sin 2 \chi \\
~\\V \approx 0 \label{spec2}
\end{array}
\end{equation}

\noindent where $c_3$ is a constant whose value depends on $\gamma$ and the
distribution of the magnetic field along the integration path.

When the medium inside the synchrotron source, or in the medium where
radiation propagates, is permeated by thermal electrons Faraday rotation
modifies the polarization charachteristics of the radiation. In fact by
Faraday effect, when the radiation crosses a region of thickness $L$
permeated by magnetic fields and thermal electrons with density $n_e$, the
angle of polarization of the radiation rotates by an angle  (see for instance
\cite{ribi})
\begin{equation}
\theta ~=~ cost ~<n_e>~L~\lambda^2 \label{Faraday}
\end{equation}

\par\noindent Inside the source this brings depolarization, because radiation
produced at different points along $L$ suffers different rotations.
Additional depolarization inside the source comes about when the magnetic
field along $L$ is not uniform because in eqs.(\ref{spec2})we have to use
$<\sin 2\chi>$ and $<\cos 2\chi>$ instead of $\sin 2\chi$ and $\cos 2\chi$.
So the degree of polarization at the source $p$ varies between 0 (random
magnetic field distribution) and:
\begin{equation}
p_{max} = \frac{3\gamma + 3}{3\gamma + 7} < 1 \label{degreep}
\end{equation}
\noindent (uniform magnetic field and Faraday rotation absent).

 Outside the source the angle of polarization rotates by Farday effect so
the Stokes Parameter of the synchrotron radiation $Q_s$ and $U_s$ measaured by
an Earth observer are different from the Stokes Parameters at the source.

A regular trend of the magnetic field inside and outside the source is
insufficient to guarantee a regular trend of the spatial distribution of
$Q_s$ and $U_s$. In fact as the line of sight moves among adjacent points on
the sky $n_e$, $<n_e>$ and $L$ fluctuate. So also the angle of Faraday
rotation $\theta$ fluctuates by a quantity
\begin{equation}
\begin{array}{c}
\delta \theta ~=~cost~\lambda^2~\delta(<n_e>L)~~~~~~~~~~~~~~~~~~~~~~~~~~~~~~~~~~~~~~~~~ \\
\\\simeq~cost~\lambda^2~\{[L~\delta(<n_e>)]^2 ~+~ [<N_e> \delta(L)]^2\}^{1/2}
~\simeq~cost~\lambda^2~L~\delta(n_e)
\end{array}
\end{equation}

\par\noindent proportional to $L$.
Because L is at least tens or hundred of pc $\delta \theta$ can be very large.
Besides this effect produced by Faraday rotation, when we look in different
directions through the interstellar medium $Q_s$ and $U_s$ may change also
because $H$, $H_p = H\sin \mu$, $\gamma$, $\beta$ and $K$ vary (see for
instance \cite{wie01}, \cite{rei01}) \footnote{e.g. $F= \left( H\sin
\mu\right)^{\frac{\gamma +1}{2}}$ is a nonzero mean, nonzero variance
variable; $\cos 2 \chi$ and $\sin 2 \chi$ are zero mean, nonzero variance
variables}.

 We conclude that angle and degree of polarization of the
synchrotron radiation fluctuates when the line of sight moves among adjacent
regions on the sky, a conclusion supported by recent observations which show
that the spatial distribution  of the polarized component of the synchrotron
radiation at intermediate and small angular scales is highly structured. On
maps we see in fact lines ({\it channels}) along which $p$ goes to zero and
sudden rotations of the plane of polarization when the line of sight goes
from a side to the other of a channel \cite{stru}.

We expect therefore that at small and intermediate angular scales $Q_s$ and
$U_s$ are stochastic functions of the direction of observations so we can
write:

\begin{equation}
\left<Q^2_s\right> = \left<U^2_s\right> \label{main}
\end{equation}

 Exceptions to this behaviour can be expected when
along the line of sight there are peculiar regions characterized by special
field configurations and/or by large densities of thermal electrons, like in
the well known loops and spurs or in HII regions, where systematic effects
overcome random effects. Nature and origin of these features, clearly visible
on Brouw and Spoelstra maps of the polarized component of the galactic
background \cite{brou}, are discussed for instance by \cite{salt}. Usually
these regions are close to the observer and have large angular dimensions, so
their effects can be removed if one evaluates the angular power spectrum of
the radiation distribution and limits his analysis to angular scales below
few degrees (multipole order $l>150$) \footnote{At larger angular scales
(smaller values of l) the number of independent samples one collects
decreases until it is insufficient to carry on statistical analysis and the
method of separation of CMB and foregrounds, we will illustrate in the
following, fails.}.

Being aware that this is a very important issue we tested its validity
analyzing the distribution of the measured values of $Q$ and $U$ on maps
extracted from the Effelsberg survey (\cite{surv} and \cite{dun99}) of the
diffuse radio emission. We found (see Appendix 2) that up to angular scales
of 5 degrees (l close to 36) the difference $<Q^2>-<U^2>$ is fully consistent
with zero.

 Because of eq. (\ref{main}) if $Q_s \ne 0$
also $U_s \ne 0$, so the synchrotron radiation has both electric and magnetic
modes:
\begin{equation} a^{E,s} \ne
0, \hskip2cm a^{B,s} \ne 0
\end{equation}

\noindent and (see eqs. (\ref{syn1}) and (\ref{main})):
\begin{equation}
\left<(a^{E,s})^2\right> = \left<(a^{B,s})^2\right> \label{syn}
\end{equation}

\section{Stokes Parameters of the CMB}

In a homogeneous and isotropic Universe only temperature and intensity $I=I_l
+ I_r$ change as the Universe expands : both decrease adiabatically. Because
this is true for $I_l$ and $I_r$ separately, we do not expect anisotropy nor
polarization therefore $Q=0$ and $U = I_u =0$ are natural consequences.

On the contrary, inhomogeneities and perturbations of matter density or of
gravitational field, induce anisotropy and polarization of the CMB. At the
recombination epoch linear polarization appears as a by product of the
Thomson scattering of the CMB on the free electrons of the primordial plasma.
The polarizarion tensor it gives can be calculated solving the Boltzman
transfer equations of the radiation in a nonstationary plasma permeated by a
variable and inhomogeneous gravitational field \cite{bas80}, \cite{saz84},
\cite{har93}, \cite{saz95}.

The gravitational field is made of a background field, with homogeneous and
isotropic FRW metric, and an inhomogeneous and variable mix of waves: density
fluctuations, velocity fluctuations, and gravitational waves. Because of
their transformation laws these waves are also said scalar, vector and tensor
perturbations, respectively.

\subsection{Scalar (density) perturbations}

Scalar (density) perturbations affect the gravitational field, the density of
matter and its velocity distribution. They were discovered studying the matter
distribution in our Universe on scales from $\sim 1$ Mpc to $\sim 100$ Mpc. It
is firmly believed they are the seeds of the large scale structure of the
Universe and are reflected by the large scale CMB anisotropy detected for the
first time at the beginning of the `90s \cite{str92}, \cite{smo92}. Their
existence is predicted by the great majority of models of the early Universe.

Observation shows that the effects of these perturbations are small, so we can
treat them as small variations $\delta_l$, $\delta_r$, $\delta_u$ of $I_l$,
$I_r$ and $I_u$.  Introducing the auxiliary functions $\alpha$ and $\beta$:
\begin{equation}
\begin{array}{c}
\delta_l + \delta_r = (\mu^2 - {1 \over 3}) \alpha, \\
~\\ \delta_l - \delta_r = ( 1 - \mu^2) \beta, \\
~\\ \delta_u =0,
\end{array}
\end{equation}

\noindent ($\mu$ is the angle between the line of sight and the wave vector)
for plane waves the Boltzman equations (see for instance \cite{saz95},
\cite{saz96a},  \cite{saz96b} and reference therein) become:

\begin{equation}
\begin{array}{c}
\frac{d \alpha}{d \eta} = F - \frac{9}{10} \sigma_T n_e a(\eta) \alpha -
\frac{6}{10} \sigma_T n_e a(\eta) \beta \\
~\\ \frac{d \beta}{d \eta} = - \frac{1}{10} \sigma_T n_e a(\eta) \alpha -
\frac{4}{10} \sigma_T n_e a(\eta) \beta
\end{array}
\end{equation}

\noindent where $F$ is the gravitational force which drives both anisotropy
and polarization, $\sigma_T$ is the Thomson cross-section, $n_e$ is the
density of free electrons, and $a(\eta)$ the scale factor.

These equations give:
\begin{equation}
\begin{array}{c}
 Q= -\frac{1}{7}(1 - \mu^2) \int F(\eta) \left( e^{-\tau} - e^{-{3 \over
10} \tau}\right) d\eta \\
~\\ U= 0
\end{array}
\end{equation}

\noindent where $\tau(\eta)$ is the optical depth of the region where the
phenomenon occurs. Rotating the coordinate system we can generate a new pair
of Stokes parameters ($Q^\prime ,U^\prime$): no matter which is the system of
reference we choose these parameters satisfy the symmetry parity condition.

Because there is always a system in which $Q \ne 0$ and $U=0$, we may conclude
that, in a system dominated by the primordial density perturbations (see, for
instance \cite{dol90}) magnetic modes of the CMB polarization vanish and only
electric modes exist \cite{sel97}. Therefore:
\begin{equation}
a^{E,d} \ne 0 \hskip2cm a^{B,d}=0 \label{aed}
\end{equation}
where index $d$ stays for {\it density perturbation}.

\subsection{Vector (velocity) perturbations}

Vector perturbations, associated to rotational effects, perturb only velocity
and gravitational field. They are not predicted by the inflation theory and
it is common believe that they do not contribute to the anisotropy and
polarization of the CMB.

\subsection{Tensor (gravitational waves) perturbations}

Gravitational waves (tensor modes) and gravitational lensing of large scale
structures of the Universe induce B modes in the distribution of CMBP(see for
instance \cite{hu})) . Gravitational lensing gives a power spectrum  at least
an order of magnitude below the power spectrum of the E mode polarized signal
produced by scalar perturbations, with a maximum  at $l\simeq 1000)$. The
power spectrum of the B mode polarization produced by gravity waves is
maximun at $l\simeq 90$, is definitely below the power spectrum produced by
gravitational lensing  for $l>100$ and only at very large angular scales
($l\simeq 20$) it may be comparable to the scalar E modes.

\par~\par If one excludes very large angular scales we may conclude that CMBP
is dominated by E-modes. B-modes are just a contamination by B-modes at
levels of $10\%$ or less.

\section{Separation of the polarized components of Synchrotron and CMB Radiation}
The CMB radiation we receive is mixed with foregrounds of local origin. When
the anisotropy of the CMB was detected, to remove the foregrounds from maps
of the diffuse radiation, data were reorganized in the following way:
\begin{equation}
\hat T_d(x,y) =  \hat T_n(x,y) + \hat M_i(x,y) \hat T_{i,c}(x,y)
\end{equation}

\noindent where $\hat T_d$, is a two dimension vector  (map) which gives the
total signal measured at different points $(x,y)$ on the sky, $\hat T_n$ the
noise vector, $\hat M$ the matrix which combines the components $\hat
T_{i,c}$ of the signal. At each point $(x,y)$ we can in fact write:
\begin{equation}
T_d = g_nT_n + g_{cmb}T_{cmb, c} + g_{syn}T_{syn, c} + g_{ff}T_{ff, c}
+g_{dust}T_{dust, c} +...
\end{equation}

\noindent where $g_i$ are weights, given by $M$, $T_{syn, c}$ is the
synchrotron component, $T_{ff, c}$ the free-free emission component,
$T_{dust, c}$ the dust contribution and so on. Using just one map the signal
components cannot be disentangled. If however one has maps of the same region
of sky made at different frequencies it is possible to write a system of
equations. Provided the number of maps and equations is sufficient, the
system can be solved and the components of $T_d$ separated, breaking the
degeneracy. We end up with a map of $T_{cmb}$ which can be used to estimate
the CMB anisotropy.

When we look for polarization at each point on the sky we measure tensors
instead of scalar quantities, therefore to disentangle the polarized
components of the CMB we need a greater number of equations. Here we will
concentrate on the separation of the two dominant components of the polarized
diffuse radiation: galactic synchrotron (plus dust) foreground and CMBP.

\subsection{The estimator $D$}
Instead of observing the same region of sky at many frequencies, we suggest a
different approach. It takes advantage of the differences between the
statistical properties of the two most important components of the polarized
diffuse radiation (CMB (background) and synchrotron (foreground) radiation)
and does not require multifrequency maps.
\par~\par

\noindent We define the estimator:
\begin{equation}
D = \left<(a^E)^2\right> - \left<(a^B)^2\right>
\end{equation}
\noindent where $a^E = a^{E,s} + a^{E,c}$ and $a^B = a^{B,s} + a^{B,c}$ (here
and in the following indexes $s$ or $c$ stay for {\it synchrotron} and CMB,
respectively).
\par \noindent Because
\begin{itemize}

\item $E$ and $B$ modes of synchrotron do not correlate each other neither
correlate with the CMB modes

\noindent  $\left<(a^E)^2\right> = \left<(a^{E,s})^2\right> +
\left<(a^{E,c})^2\right> + 2\left<a^{E,s} a^{E,c}\right> =
\left<(a^{E,s})^2\right> + \left<(a^{E,c})^2\right>$,

\noindent $\left<(a^B)^2\right> = \left<(a^{B,s})^2\right> +
\left<(a^{B,c})^2\right> + 2\left<a^{B,s} a^{B,c}\right> =
\left<(a^{B,s})^2\right> + \left<(a^{B,c})^2\right>$,

\item $~~~~~~~~~~~~~<a^{E,s}> = <a^{B,s}> \neq 0$

\item
$$
a^{E,c} = a^{E,d} \ne 0 \hskip4cm a^{B,c} = a^{B,t} \leq 0.1~a^{E,d}
$$
\end{itemize}
\par\noindent where indexes $d$ and $t$ stay for $density$ (or $scalar$) and $tensor$
perturbation, respectively.

\par~\par\noindent $D$ gives an estimate of the E-mode excess in maps of
the polarized diffuse radiation. If tensor perturbations are negligible this
excess is the CMBP signal. If tensor perturbations are important the excess is
a lower limit with a systematic difference from the true value which in the
worst condition (maximum contribution to CMBP from gravitational waves and
gravitational lensing) reaches a maximum value of $10\%$.

Let` s now consider the angular power spectrum of $D$. For multipole $l$ we
can write:
\begin{equation}
D_l = (a^E_l)^2 - (a^B_l)^2 = \frac{1}{2l + 1} \sum_{m = -l}^l \left(|a^E_{l
m}|^2 - |a^B_{l m}|^2 \right) \label{dest}
\end{equation}

\noindent where $|a^E_{l m}|^2$ and $|a^B_{l m}|^2$ are random variables with
gaussian distribution $p(a_{lm}^{E,B})$ (see eqs.(\ref{gauss1}) and
(\ref{gauss2}) in Appendix A). According the ergodic theorem (in the limit of
infinite maps, the average over 2D space is equivalent to the average over
realisations) the average value of $D_l$ is equal to the difference of the
average values of $|a^E_{l m}|^2$ and $|a^B_{l m}|^2$ summed over $m$. Taking
into account equation (\ref{gauss2}) we can therefore write:
\begin{equation}
\left< D_l \right> =  \left(|a^E_{l}|^2 -  |a^B_{l}|^2 \right) \label{D1}
\end{equation}

\noindent where \footnote {Equation (\ref{gauss0}) is an explicit form of the
average of the stochastic variables $|a_{lm}^E|^2$ and $|a_{lm}^B|^2$ over a
probability density $p(a_{lm}^{E,B})$, the short form being triangle
brackets}:

\begin{equation} \begin{array}{c}
(a^E_l)^2 = \int\limits^{\infty}_{-\infty}|a_{lm}^E|^2 p(a_{lm}^E) d^2
a^E_{lm} = \left< |a_{lm}^E|^2\right>
  \\
~\\
(a^B_l)^2 = \int\limits^{\infty}_{-\infty}|a_{lm}^B|^2 p(a_{lm}^B) d^2
a^B_{lm} = \left< |a_{lm}^B|^2\right> \label{coco}
\end{array}
\label{gauss0}
\end{equation}

\noindent Comparing eq.(\ref{coco}) with the ordinary definition of multipole
coefficients:
\begin{equation}
C_l^{E,B}= \frac{1}{2l + 1} \sum_{ m= -l}^{ l} \left< |a_{lm}^{E,B}|^2\right>
\label{multipole}
\end{equation}
\noindent we can write:
\begin{equation} \begin{array}{c}
 (a^E_l)^2 = C^E_l \\
~\\
(a^B_l)^2 = C^B_l
 \label{equal}
\end{array}
\end{equation}

\subsection{Separation uncertainty}
For synchrotron radiation $\left< D_l \right>$ should be zero, non zero for
density perturbations, but on real maps it is always different from zero. In
fact a map is just a realization of a stochastic process and the amplitudes
of  $|a^E_{l m}|^2$ and $|a^B_{l m}|^2$, averaged over 2D sphere, have
uncertainties which add quadratically, so $\left< D_l \right> \ne 0$ even in
the case of synchrotron polarization. This effect, very similar to the well
known {\it cosmic variance} of anisotropy   \cite{var}, \cite{saz95b} (the
real Universe is just a realisation of a stochastic process, therefore there
will be always a difference between the realization we measure and the
expectation value) does not vanish if observations are repeated.

\par~\par The variance of $D_l$ is
\begin{equation}
{\bf V} (D_l^2) = \left< D_l^2 \right> - \left< D_l \right>^2 \label{var0}
\end{equation}

The quantities $a^E$ and $a^B$, being sums over $m$ of $2l+1$ stochastic
values with gaussian distribution, have a $\chi^2$ distribution with $2*(2l
+1)$ degrees of freedom, so their variance is

\begin{equation}
\delta (a^{E, B})^2 \propto \sqrt{\frac{2}{2 l + 1}} \label{var1}
\end{equation}

\noindent More explicitly
\begin{equation}
\left< D^2_l \right> = \frac{1}{2 l +1}  \left((a_l^E)^4 +  (a_l^B)^4 \right)
\label {Dl20}
\end{equation}
\noindent When the synchrotron foreground is dominant ($a_l^E=a_l^B$)
\begin{equation}
\left< D^2_l \right> = \frac{2}{2 l +1} (a^E_l)^4 \label{Dl21}
\end{equation}

\noindent in agreement with (\ref{var1}).

\subsection{A criterium for CMBP detection}
The synchrotron foreground is a sort of {\it system noise} which hampers the
detection of the {\it signal}, the CMB polarization. At frequencies
sufficiently high (above $\sim 50$ GHz, see next section) the noise is small
compared to the signal therefore direct detection of CMBP is possible. At low
frequencies on the contrary the CMBP signal is buried in the noise created by
synchrotron and dust emission. In this case to recognize the presence of the
CMB polarization we can use our estimator $D$.

At angular scale $l$, to be detectable the CMBP must satisfy the condition

\begin{equation}
C_l^{E,c} > A \cdot C_l^{E,s} \label{sn1}
\end{equation}

\noindent where $C_l^{E,i}$ are the coefficients of the multipole expansion of
the $E$ modes and $A$ is the confindence level of the signal detectability.
In a similar way we can write for our estimator:
\begin{equation}
D_l^{E,c} > A \cdot D_l^{E,s} \label{sn2}
\end{equation}
\noindent where
\begin{equation}
D_l^{E,s} = \sqrt{\frac{2}{2l+1}} (a_l^{E,s})^2 \label{sn3}
\end{equation}

\par\noindent If one neglects tensor perturbations ($D_l^{E,c} = C_l^{E,d}$)
the criterium for the CMBP detection becomes
\begin{equation}
D_l^{E,c} \geq A \cdot \sqrt{\frac{2}{2l+1}} C_l^{E,s} \label{sn4}
\end{equation}

\noindent so by $D$ we get the CMBP level with an uncertainty ~$\simeq
\sqrt{\frac{2}{2l+1}} C_l^{E,s}$ which decreases as $l$ and the angular
resolution increase. Tensor perturbations, if present, add to this
uncertainty a systematic uncertainty $\leq 10\%$.

\section{Angular power spectra of polarized synchrotron.}

The angular power spectra of the polarized component of the synchrotron
radiation have been studied by \cite{bur01}, \cite{bur02}, \cite{tuc00},
\cite{tuc02}, \cite{brus02} and \cite{gia02} using the few partial maps of
the polarized diffuse radiation available in literature.
 It appears that the power spectra of the degree of polarization $p$ and of electric and
magnetic modes $E$ and $B$ follow power laws of $l$ up to $l \sim 10^3$. For
the degree of polarization the spectral index is $\alpha_p \simeq 1.6 - 1.8$.
For the $E$ and $B$ modes different authors get different values of the
spectral index.

According the authors of papers \cite{tuc02} and \cite{brus02}, Parkes data
give:

\begin{equation}
C^{E,B}_l \sim C_{0, E, B} \left(\frac{l_0}{l}\right)^{\alpha_{E,B}}
\label{syn111}
\end{equation}

\noindent  with $\alpha_E \approx \alpha_B \approx 1.4 \div 1.5$ and
dependence of $\alpha$ on the region of sky and the frequency.

In paper \cite{bur01}, using Effelsberg and Parkes data, the authors get:
\begin{equation}
C^{E,B}_l = C_0 \cdot 10^{-10}  \left(\frac{450}{l}\right)^{\alpha} \cdot
\left(\frac{2.4 \mbox{GHz}}{\nu}\right)^{2\beta} \label{synEBB}
\end{equation}
with  $\alpha =(1.8 \pm 0.3)$, $\beta = 2.9$ and $C_0= (1.6 \pm 1)$ (here we
adjusted the original expression given in \cite{bur01} writing it in
adimensional form).

In paper \cite{gia02} the authors, using a completely different set of
observational data (\cite{jon98} and \cite{gia01}) conclude that in the
multipole range $l=40 - 250$ the spectral indexes of the $E$ and $B$ modes
are $\alpha_E = (1.57 \pm 012)$ and $\alpha_B = (1.45 \pm 0.12)$
respectively, while for the polarized intensity the spectral index is
$\alpha_I = (2.37 \pm 0.21)$.

Inside the multipole range $(10^2 \le l \le 10^3)$ the values of
$\alpha_{E,B}$ obtained by the three groups are marginally consistent and
reasonably close to $\sim 1.8$. At lower value of $l$ definite differences
exist between the results obtained by \cite{bur01} and by \cite{gia02}: these
differences must be understood, but they are not important here because our
statistical method cannot be used for small values of $l$ when the number of
samples becomes insufficient to carry on statistical analyses. For high values
of $l$ , if one excludes extrapolations and models (eg. \cite{gia02}), there
are no data in literature, but this is not a limitation because for large
values of $l$ the results of our method are practically independent from the
spectral index.

All the above authors got their results analyzing low frequency data (1.4, 2.4
and 2.7 GHz), therefore the extension of their spectra to tens of GHz, the
region where CMB observations are usually made, depend on the accuracy of
$\beta$, the temperature spectral index of the galactic synchrotron
radiation. A common choice is $\beta=2.9$ but in literature there are values
of $\beta$ ranging between $\approx 2.5$ and $\approx 3.5$. Moreover $\beta$
depends on the frequency and the region of sky where measurements are made
(see \cite{ele1}, \cite{salt}, \cite{zanno}, \cite{bersa}). Last but not least
many of the values of $\beta$ in literature have been obtained measuring  the
total (polarized plus unpolarized) galactic emission. In absence of Faraday
effect $\beta , \beta_{pol}$ and $\beta_{unpol}$, the spectral indexes  of
the total, polarized and unpolarized components of the galactic emission,
should be identical (see eqs. (\ref{spec1}),(\ref{spec2})). However when
Faraday effect with its $\nu^{-2}$ frequency dependence is present, we expect
that, as frequency increases, the measured value of the degree of
polarization (se eq. (\ref{degreep})) increases, up to
$$
p \leq p_{max} = \frac{3\beta -3}{3\beta -1}
$$

\noindent therefore we should observe $\beta_{pol} \leq \beta$. The expected
differences are however well inside the error bars of the data in literature
so at present we can neglect them and assume $\beta \simeq \beta_{pol} \simeq
\beta_{unpol}$.

Instead of extrapolating low frequency results it would be better to look for
direct observations of the galactic emission and its polarized component at
higher frequencies. Unfortunately above 5 GHz observations of the galactic
synchrotron spectrum and its distribution are rare and incomplete.
  At 33 GHz observations by \cite{dav99} give at some patches on the sky
a galactic temperature of about $2 \div 4 \mu K$ from which follows that at
the same frequency we can expect polarized foreground signals up to several
$\mu K$. At 14.5 GHz observations made at OVRO \cite{muk02} give synchrotron
signals of 175 $\mu K$, equivalent to about 15 $\mu K$ at 33 GHz of which up
to 10 $\mu K$ can be polarized.

In conclusion there are large uncertainties on the frequency above which
observations of the CMBP are practically unaffected by the polarized
component of the galactic diffuse emission. To be on the safe side we can set
it at 50 GHz (see for instance \cite{bur01}, \cite{bur02}, \cite{tuc00},
\cite{tuc02} and \cite{brus02}). Above 50 GHz  CMBP definitely overcomes the
polarized synchrotron foreground. Below 50 GHz contamination by the galactic
emission can be important but its evaluation, usually made by multifrequency
observations, is dubious.

\begin{figure}
\vspace{-15pt} \centerline{\hspace{30pt}\epsfig{file=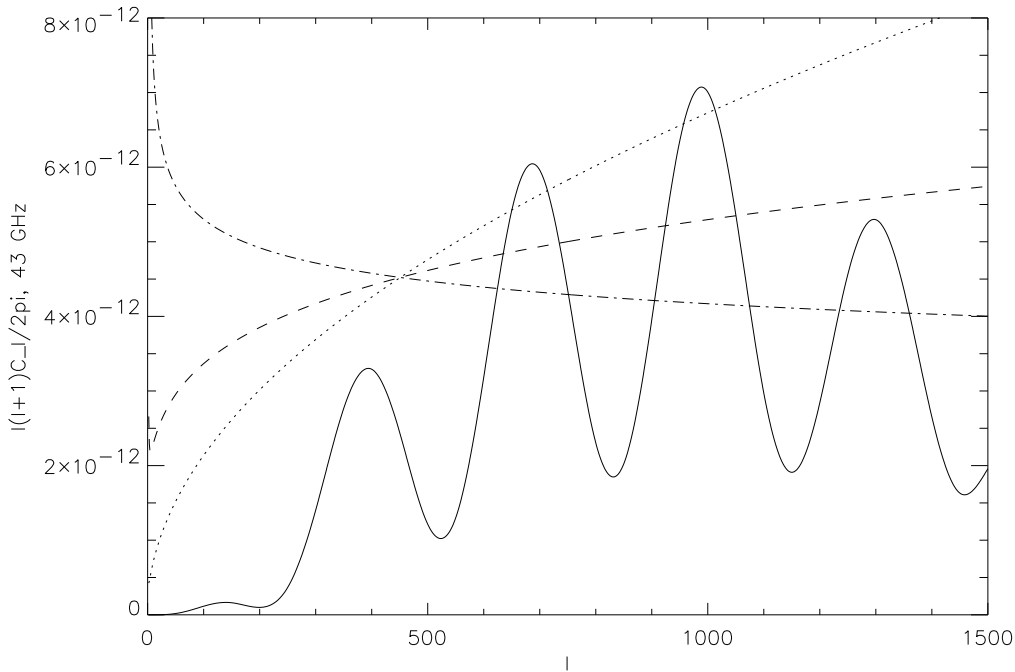,height=3.5in}}
\vspace{-12pt} \caption{\small Power spectrum versus multipole order $l$ of
the polarized components of CMB (solid line) and three possible power spectra
of the galactic synchrotron radiation ($\alpha = 1.5$ (dotted line), $\alpha
= 1.8$ (dashed line) and $\alpha = 2.1$ (dot-dash line)) calculated at 43
GHz. (See text for details of calculations and model)} \label{f9}
\end{figure}

This conclusion is supported by figure \ref{f9} where we plotted,  versus the
multipole order $l$, the power spectra of CMBP and galactic synchrotron at 43
GHz. The CMB spectrum has been calculated by CMBFAST \cite{zal02} assuming
standard cosmological conditions (CMB power spectrum normalized to the COBE
data at low $l$, $\Omega_b = 0.05$, $\Omega_{CDM}=0.3$ , $\Omega_{\Lambda} =
0.65$, $\Omega_{\nu} =0$, $H_0 =65$ km/sec/Mpc, $T_{CMB}=2.726 K$ , $Y_{He}
=0.24$, standard recombination). The synchrotron spectrum has been calculated
assuming for $E_l$ and $B_l$ the scaling law (\ref{synEBB}) with $C_0=2.6$
(most pessimistic case), $\alpha = 1.5;~ 1.8;~ 2.1$ and $\beta =2.9$
respectively. It appears that at 43 GHz the CMB power is comparable to the
synchrotron power only at very small angular scales ($l \geq 500$).

 Similar calculations at other frequencies
confirm that only above $\simeq 50$ GHz and at small angular scales (large
values of $l$) the CMBP power spectrum overcomes the synchrotron spectrum.
Below $\simeq 50$ GHz direct detection of the CMBP is almost impossible even
at small angular scale.
\par~\par
To overcome this limit we can use our estimator. To see how it improves the
CMBP detectability we calculated the angular power spectrum of $D^s_l$, the
estimator we expect when the diffuse radiation is dominated by synchrotron
radiation. From eqs.(\ref{multipole}), (\ref{Dl21}) and (\ref{synEBB}) we get:

\begin{equation}
\sqrt{\left< (D^s_l)^2 \right>} =C_0 \cdot 10^{-10} \sqrt{\frac{2}{2l +1}}
\left(\frac{450}{l}\right)^{\alpha} \cdot  \left(\frac{2.4
\mbox{GHz}}{\nu}\right)^{2 \beta} \label{synD}
\end{equation}

\noindent a quantity which can be directly compared with the power spectrum of
CMBP. In fact eqs.(\ref{aed}), (\ref{D1}), (\ref{equal}) and (\ref{Dl20})
show that the power spectrum $D^d_l$ of the estimator evaluated when the sky
is dominated by CMBP density perturbations coincides with the power spectrum
of CMBP produced by density perturbations.

\begin{figure}
\epsfig{file=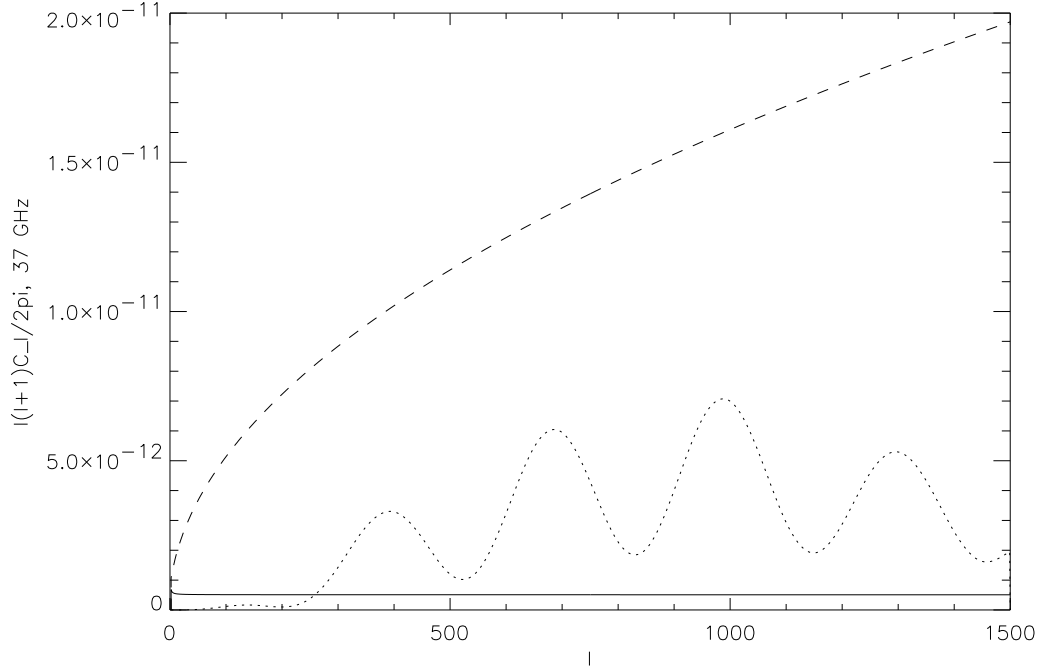,height=3.5in} \caption{\small Power spectrum at 37 GHz
of the expected value of the estimator  for a galactic synchrotron dominated
sky (solid line). The dotted line represents both the estimator and the CMBP
power spectrum for a CMBP dominated sky. The dashed line is the power
spectrum  of the polarized component of the galactic synchrotron radiation
(see text for details of model and calculations)}
 \label{fD}
\end{figure}

Figure \ref{fD} shows at 37 GHz the power spectra at 37 GHz of: i)$D^s_l$, the
estimator for a sky dominated by galactic synchrotron  (solid line ,
calculated using eq.(\ref{synEBB})), ii)$D^d_l$, the estimator for a sky
dominated by CMBP. It coincides with the power spectrum of CMBP (dotted line,
calculated as in figure \ref{f9} with CMBFAST using the same standard
cosmological conditions), iii)the power spectrum of the polarized component of
the galactic synchrotron radiation (dashed line, calculated using
eq.(\ref{synEBB})). As expected at 37 GHz the CMBP power is well below the
synchrotron power, therefore direct observations of CMBP are impossible (the
maximum value of the CMBP power is about 2.5 times below the synchrotron
power at the same $l$). However above $l\simeq 250$ the power of the
synchrotron estimator is definitely below the power of the CMBP estimator: at
$l = 1000$ the ratio CMBP/D is maximum and close to $7$. This confirms that
the use of $D$ allows to recognize the CMBP also at frequencies well below 50
GHz.

\section{Simulations}

To further test the capability of our estimator we studied the separation of
CMB and galactic synchrotron using measured instead of expected values of
$D_l$.The measurements were simulated by random numbers, with gaussian
distribution $\sim N(0,1)$, zero mean and unity variance (see eqs
(\ref{gauss1}) - (\ref{gauss2})). Two series of  $2l+1$ random numbers gave
representations of  $a^E_{lm}$ and $a^B_{lm}$ respectively and from them we
obtained $D_l$ (see eq.(\ref{dest})).

\begin{figure}
\vspace{-15pt} \centerline{\hspace{30pt}\epsfig{file=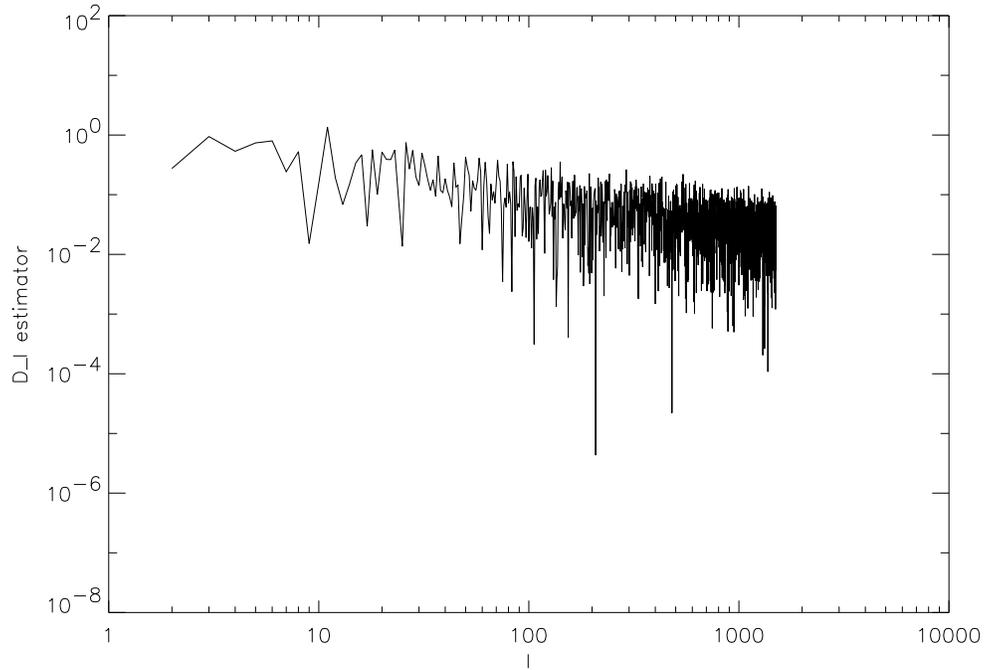,height=3.5in}}
\vspace{-12pt} \caption{\small Multipole power spectrum of simulated
measurements of the estimator $D$ for a synchrotron dominated sky (see text)
with infinite angular resolution (no smoothing on $l$, ($\Delta l = 1$))}
\label{f6}
\end{figure}
Then, to take into account that real data are collected with a finite angular
resolution (e.g. Boomerang data come from regions whose angular extension is
equivalent to $\Delta l \sim 100$ \cite{boo00}, \cite{boo01a}, \cite{boo01b})
we averaged the above values of $D_l$ on intervals $\Delta l = n$:
\begin{equation}
 \hat D_l = \frac{1}{n+1} |\sum\limits_{k = 0}^{k = +n} D_{l-n/2+k}|
\end{equation}

\begin{figure}
\vspace{-15pt} \centerline{\hspace{30pt}\epsfig{file=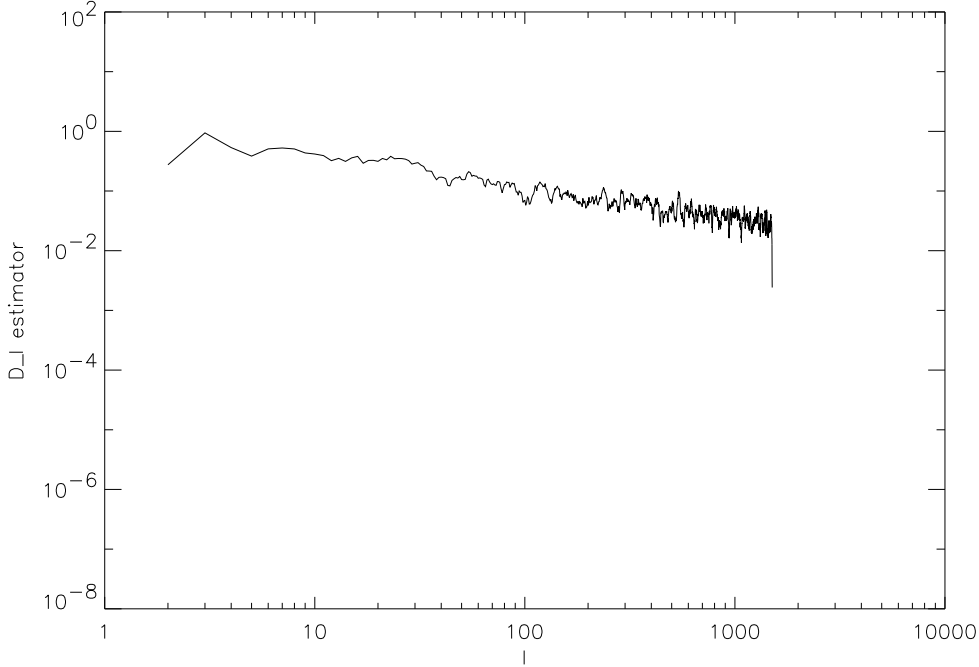,height=3.5in}}
\vspace{-12pt} \caption{\small Same as figure \ref{f6} with finite angular
resolution (smoothing on $\Delta l = 10$)} \label{f7}
\end{figure}

\begin{figure}
\vspace{-15pt} \centerline{\hspace{30pt}\epsfig{file=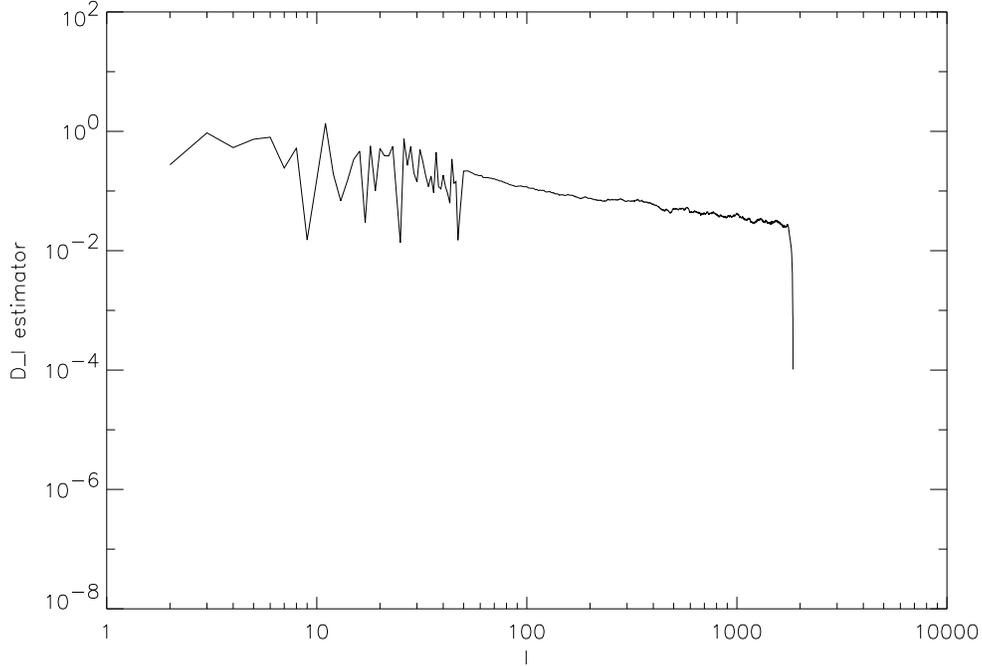,height=3.5in}}
\vspace{-12pt} \caption{\small Same as figure \ref{f6} with finite angular
resolution (smoothing on $\Delta l = 100$)} \label{f8}
\end{figure}

\noindent Figure \ref{f6},  figure \ref{f7} and figure \ref{f8} show $ |\hat
D_l|$ (the sign of $D_l$ is arbitrary) versus $l$, for $\Delta l =1$, $\Delta
l =10$ and $\Delta l = 100$ respectively: the very large fluctuations of the
estimator are drastically reduced as soon as $\Delta l$ increases.

\begin{figure}
\vspace{-15pt} \centerline{\hspace{30pt}\epsfig{file=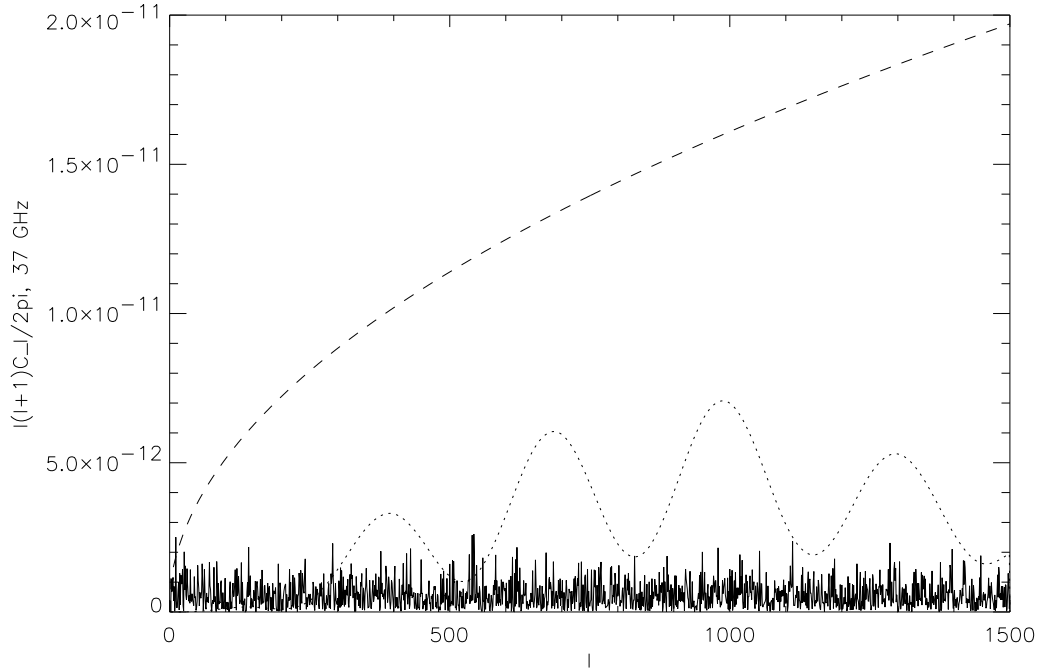,height=3.5in}}
\vspace{-12pt} \caption{\small Similar to figure \ref{fD}. Here we plot
simulated measurements with infinite angular resolution (no smoothing on $l$)
instead of the expectation value of the estimator at 37 GHz for a synchrotron
dominated sky. The dotted line give both the estimator and the CMBP power
spectrum for a CMBP dominated sky. The dashed line is the power spectrum of
the galacic synchrotron when its expected contribution is maximum (eq.
(\ref{synEBB}) with $\beta = 2.9$, $\alpha = 1.5$ and $C_o = 2.6$) (see text
for details)} \label{f10}
\end{figure}

\begin{figure}
\vspace{-15pt} \centerline{\hspace{30pt}\epsfig{file=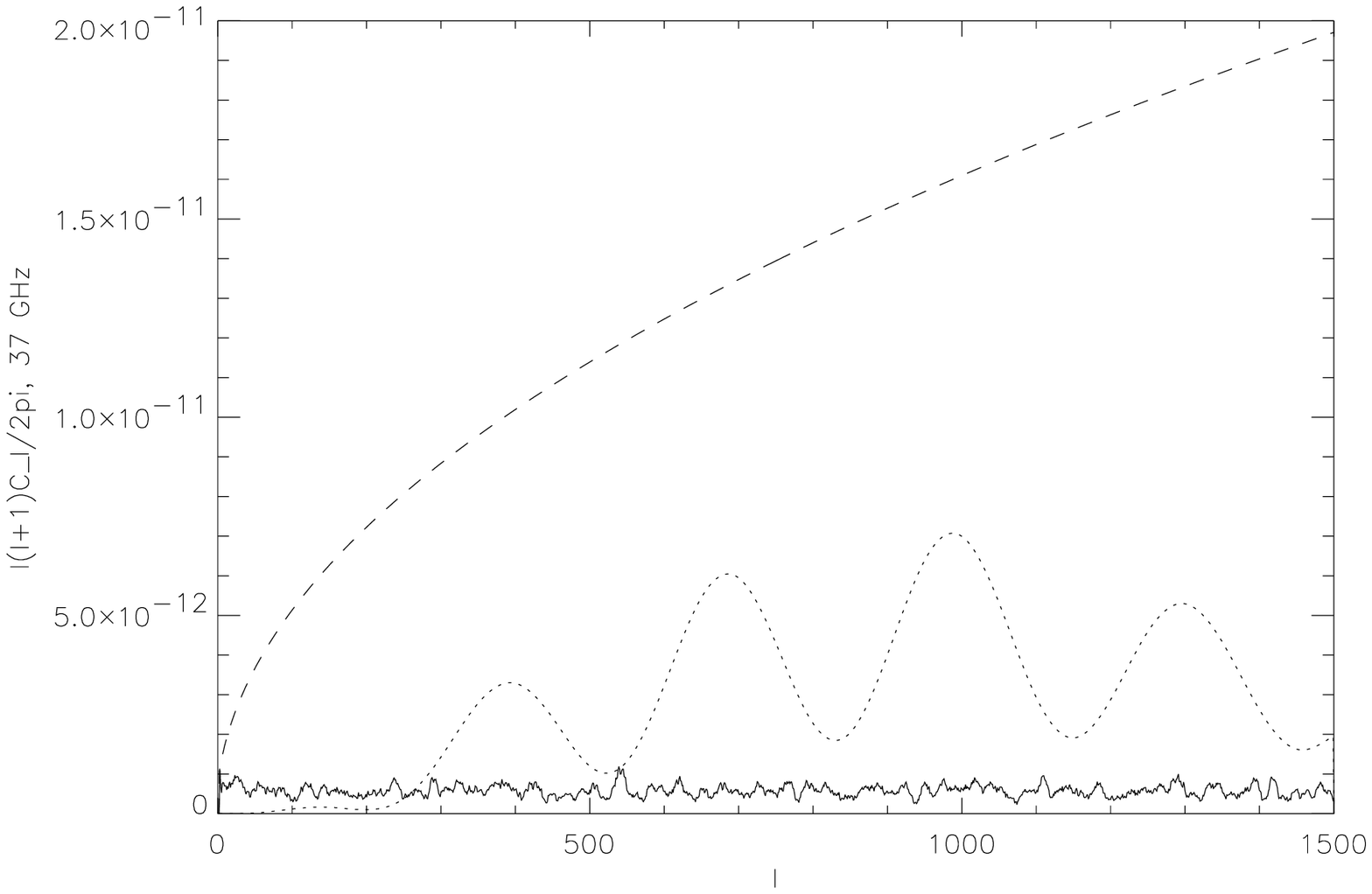,height=3.5in}}
\vspace{-12pt} \caption{\small Same as figure \ref{f10} with finite angular
resolution (smoothing on $\Delta l = 10$)} \label{f11}
\end{figure}

\begin{figure}
\vspace{-15pt} \centerline{\hspace{30pt}\epsfig{file=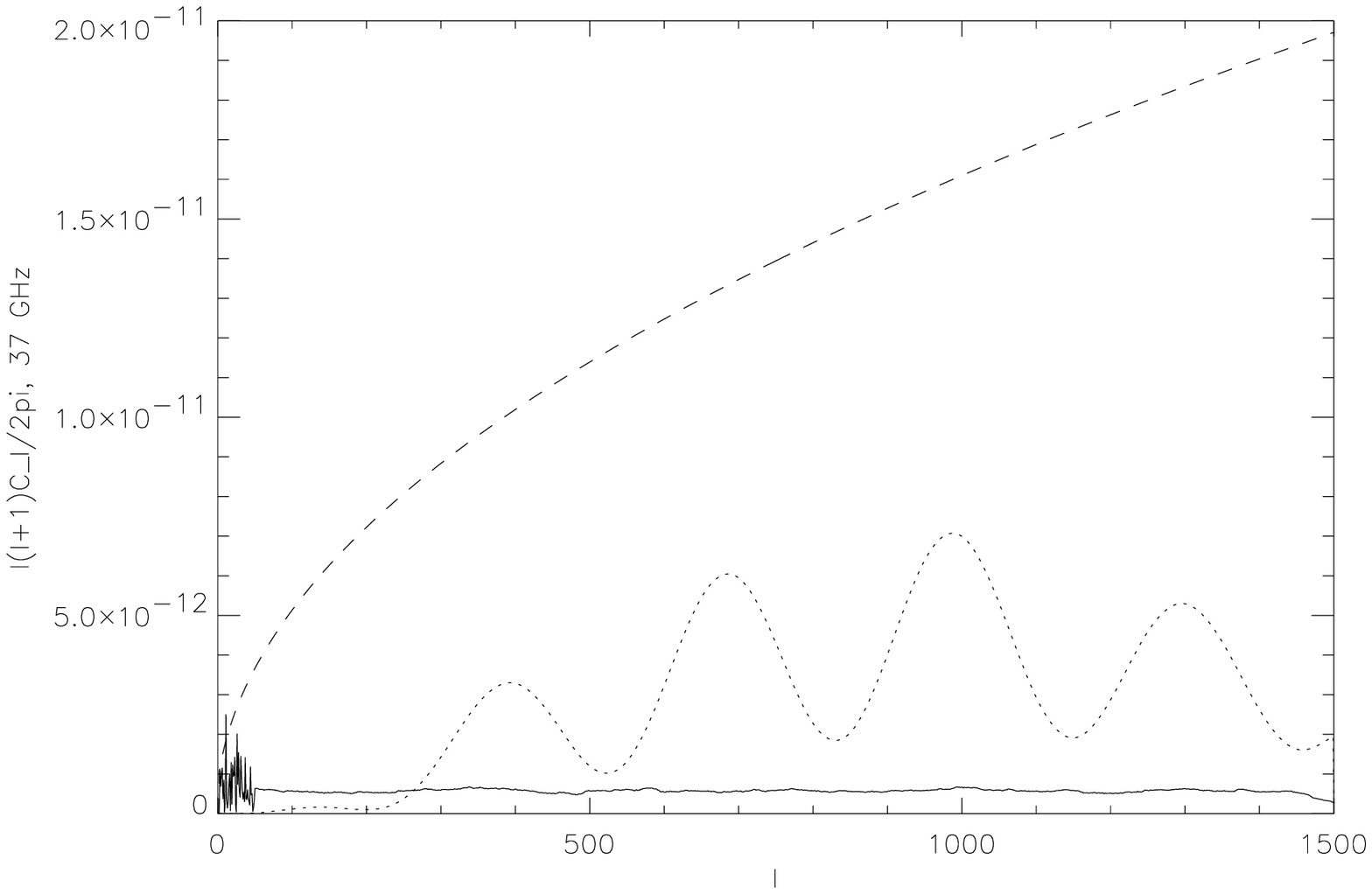,height=3.5in}}
\vspace{-12pt} \caption{\small Same as figure \ref{f10} with finite angular
resolution (smoothing on $\Delta l = 100$)} \label{f12}
\end{figure}

Figure \ref{f10}, figure \ref{f11} and figure \ref{f12} are similar to figure
\ref{fD}. Here, instead of the expectation value, we plot simulated
measurements  of $D^s_l$ at 37 GHz, for $\Delta l =1$ (figure \ref{f10}),
$\Delta l =10$ (figure \ref{f11}) and $\Delta l = 100$ (figure \ref{f12}),
respectively. Once again the CMBP power spectrum comes from CMBFAST assuming
the same cosmological conditions we assumed above. The synchrotron power
spectrum is obtained from eq.(\ref{synEBB}) with $C_o=2.6, \alpha=1.5,
\beta=2.9$ (most pessimistic condition).

\begin{figure}
\vspace{-15pt} \centerline{\hspace{30pt}\epsfig{file=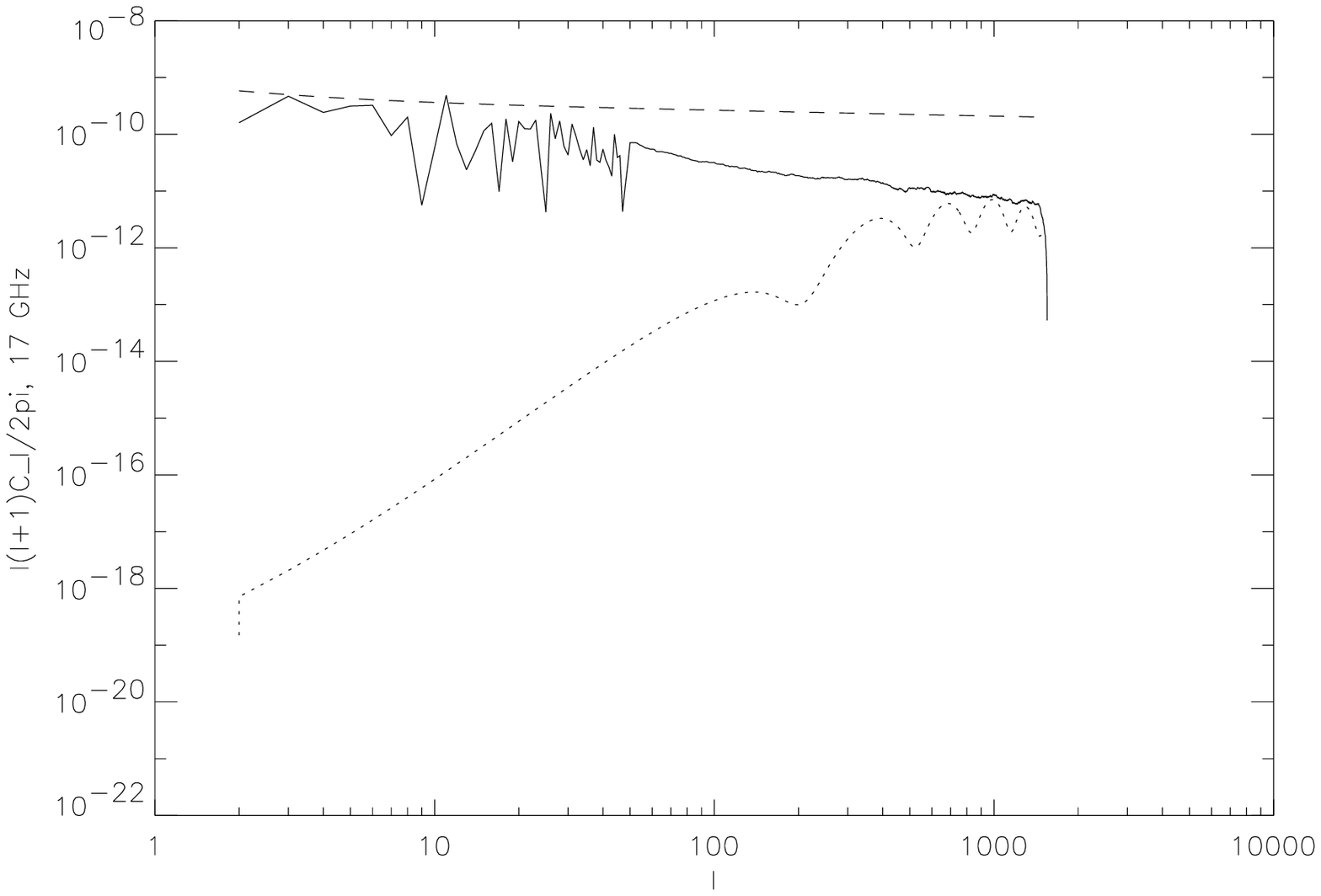,height=3.5in}}
\vspace{-12pt} \caption{\small Same as figure \ref{f12} at 17 GHz using
logarithmic scale on the vertical axis} \label{f13}
\end{figure}

Figure \ref{f13} shows at 17 GHz the same quantities we plotted in figure
\ref{f12}. For a better appreciation of the differences among the three
curves, on the vertical axis here we use a logarithmic scale. The power
spectrum of the estimator now almost touches the two highest peaks of the CMBP
spectrum. Probably 17 GHZ is the lowest frequency at which, in the most
favorable conditions, one can use $D_l$. In the most pessimistic case
($C_0=2.6$ and $\alpha=1.5$) the corresponding frequency is 25 GHz.

\section{Impact of real world experimental conditions}
We analyzed the impact of the real world experimental conditions on our method
for disentangling CMBP and foregrounds in maps of the polarized diffuse
radiation.

A polarimeter is a two channel system which (by hardware and/or software
methods) splits the sky signal in two polarized components, send them to
separate channels where they are amplified and then looks for correlations
between the two components. The system outputs are proportional to a pair of
Stokes parameters (e.g. $U$ and $Q$), or to the electric and magnetic modes
$a^E$ and $a^B$ or to a combination of them, e.g. our estimator $D$.
Unfortunately as the signals propagate through the system, noise is added to
them so at the system output the signal is mixed to noise. Moreover if the
channels are asymmetric or there are cross talks between them, the outputs
contain additional signals which simulate spurious polarization, usually an
offset of the system output from the level one should expect with an ideal
system when polarization is absent (for a discussion see for instance
\cite{spi02}). So in the real world the signal we are looking for,
$S_{true}$, is accompanied by uncorrelated noise and offset. What one measures
is
\begin{equation}
       S_{meas} = S_{true} + \eta \frac{\sigma_{noise}(t=t_o)}{\sqrt{t/t_o}} + \epsilon (S_{true} +
       N)
\label{rex} \end{equation} \noindent where $\eta \sim 1$, $\epsilon$ is the
cross talk coefficient, $N$ the noise signal and $\sigma_{noise}$ the noise
standard deviation. In commercial systems $\epsilon \sim (10^{-2} - 10^{-3})$
while in dedicated CMBP experiments values of $\epsilon < 10^{-5}$ have been
obtained \cite{cor99}. By accurate choice of the system components we can
therefore minimize the offset and further reduce it by phase modulation
techniques (see e.g. \cite{spi02}).

    At this point we have noisy maps we can use to evaluate $D$ and its power
spectrum instead of $U$ and $Q$ or $a^E$ and $a^B$. As shown by the above
simulation the signal/noise ratio for $D$ is definitely better than the
signal/noise ratio for the Stokes Parameters or $a^E$ or $a^b$ and brings it
to values similar to the ones we find when studying weak radiosources or, in
the worst situation, the CMB anisotropy. From here we can therefore go on and
extract $D$ from the remaining noise using the well known methods of  time
integration, commonly used in radioastronomy.

\section{Conclusions}
Observations of the CMB polarization are hampered by the presence of a
foreground, the polarized component of the galactic synchrotron radiation.
Only above $\sim 50 GHz$ the cosmic signal definitely overcomes the galactic
synchrotron and direct measurements of the CMBP are possible. Between $\sim
30$ GHz and $\sim 50$ GHz background and foreground are comparable. Below
$\sim 30$ GHz the polarized sky is dominated by the galactic signal.

So when measurements are made at ground observatories where atmospheric
absorption prevent observations above $\sim 40$ GHz, all programs for CMBP
measurements must include observing and analysis strategies for disentangling
the galactic synchrotron signals from the CMBP signals. A common approach is
fitting models of the intensity, frequency dependence and spatial
distribution of the cosmic and galactic signals to multifrequency maps of the
polarized diffuse radiation. Or, when observation are made at one frequency,
looking for additional data in literature, but here the accuracy of the
available data is insufficient to get firm evaluations of CMBP.

In this paper we presented a different approach which at small angular scales
($\leq 0.7^o$ ($l\geq 250$)) and down to frequencies as low as $\sim 25 GHz$
($\sim 17 GHz$ in the most favorable conditions) allows to extract the CMBP
signal from single frequency maps of the polarized diffuse radiation. It takes
advantage of the different statistical properties of the spatial distributions
of CMBP and polarized galactic synchrotron. By our estimator $D$, which
evaluates the difference between E- and B-modes, we get the polarized
component of CMB with a maximum systematic (underestimate) uncertainty of
$10\%$. This uncertainty is set by the contamination by the tensor
perturbations which add B-modes to a CMBP map dominated by the E-modes
generated by scalar(density) perturbations. Improving our knowledge of the
tensor perturbations we will reduce the above uncertainty and improve the
accuracy of our method of measuring CMBP.

The accuracy we can get with our method is the maximun one can obtain at
ground observatories  with today (2nd generation) systems for measuring the
CMB polarization. These 2nd generation experiments are just arrived on the
verge of detecting the polarized signals produced by density perturbations
(see for instance \cite{dasi} and \cite{map1}).Direct observations of the
signals associated to tensor perturbations requires new, 3rd generation,
intrinsically able to reject the foreground signals which, at present, are
still in preparation (see for instance \cite{picci}).

\section*{Acknowledgments}
We are indebted to S.Cortiglioni, E.Caretti, E.Vinjakin, J.Kaplan and J.
Delabrouille for helpful discussion. MVS acknowledges the Osservatorio
Astronomico di Capodimonte, INAF, for hospitality during preparation of this
paper.

\section*{Appendix 1: Stochastic properties of the
harmonics amplitudes}

 Here and overall in paper we suppose that $a_{lm}^s$
are complex random variables which satisfy the probability distribution law:
\begin{equation} \begin{array}{c}
p(a_{lm}^E) = \frac{1}{\pi E_l^2} exp\left( -\frac{|a_{lm}^E|^2}{
E^2_{lm}}\right) \\
~\\ p(a_{lm}^B) = \frac{1}{\pi B_l^2} exp\left( -\frac{|a_{lm}^B|^2}{
B^2_{lm}}\right)
\end{array} \end{equation}

\noindent with variance $\left< |a^E_{lm}|^2\right> =E^2_l$ and
$\left<|a^B_{lm}|^2\right> =B^2_l$.

They have all the propertiers of gaussian variables (below we omitt indexes
$E$ and $B$ in first and second equations):

\begin{equation} \begin{array}{c}
\int\limits_{-\infty}^{\infty} p(a_{lm}) d^2 a_{lm} =1 \\
~\\ \int\limits^{\infty}_{-\infty}a_{lm} p(a_{lm}) d^2 a_{lm} =0
\end{array}
\label{gauss1}
\end{equation}

\begin{equation} \begin{array}{c}
\int\limits^{\infty}_{-\infty}|a_{lm}^E|^2 p(a_{lm}^E) d^2 a^E_{lm} =E^2_l \\
~\\ \int\limits^{\infty}_{-\infty}|a_{lm}^B|^2 p(a_{lm}^B) d^2 a^B_{lm} =B^2_l
\end{array}
\label{gauss2}
\end{equation}

Setting
\begin{equation}
\left< F\right> =\int\limits^{\infty}_{-\infty}F p(a_{lm}) d^2 a_{lm}
\end{equation}

\noindent with current index $E$ or $B$, it immediately follows:
\begin{equation} \begin{array}{c}
\left<|a_{lm}^E|^4 \right> = \int\limits^{\infty}_{-\infty}|a_{lm}^E|^4
p(a_{lm}^E)
d^2 a^E_{lm} =2 E^4_l \\
~\\ \left<|a_{lm}^B|^4 \right> = \int\limits^{\infty}_{-\infty}|a_{lm}^B|^4
p(a_{lm}^B) d^2 a^B_{lm} =2 B^4_l
\end{array} \end{equation}
\par~\par~\par

\section*{Appendix 2: Stochastic properties of
the synchrotron radiation at low galactic latitudes}

In the our paper we assert that for the galactic synchrotron emission the
measured values of the Stokes Parameters $Q$ and $U$ behave as stochastic
variables and random fields with gaussian distribution. This statement
certainly holds at high galactic latitudes. At low galactic latitudes, where
we observe large scale galactic structures, regular magnetic fields
\cite{dun99} and quasi-periodic structures with typical sizes of about 250
pc, (the amplitudes of regular and irregular components are approximately
equal) this assumption has to be checked.

To do it we analyzed the distribution of the measured values of $Q$ and $U$
on regions of different extension extracted from the Effelsberg maps of the
polarized diffuse radiation \cite{surv}. The six fields were chosen within
$\pm 5^{\circ}$ from the galactic plane, at galactic longitudes between
$68^{\circ}$ and $16^{\circ}$. They had dimensions of $0.5^{\circ} \times
0.5^{\circ}$, $1^{\circ} \times 1^{\circ}$, $2^{\circ} \times 2^{\circ}$,
$3^{\circ} \times 3^{\circ}$, $4^{\circ} \times 4^{\circ}$, $5^{\circ} \times
5^{\circ}$, respectively

\begin{table}
\begin{center}
\caption{ Characteristics of the observed distributions of $Q$ and $U$
measured at low galactic latitudes in areas of various extensions (the larger
is the number of pixels the closer is the histogramm to a gaussian shape)}
\begin{tabular}{|l|c|c|c|c|c|}
\hline
size       & average       & variance & average  &  variance  & total \\
of         & value         & ~~of     & value    &  ~~of      & number\\
patch      & of $Q$        & ~~$Q$    &of $U$    &  ~~$U$     & of pixels
 \\
\hline
$0.5^{\circ} \times 0.5^{\circ}$    &16.2   &564.0  &16.9  &733.4 &256 \\
\hline
$1^{\circ} \times 1^{\circ}$        &6.2   &269.3  &-6.2  &311.9  &961\\
\hline
$2^{\circ} \times 2^{\circ}$        &0.07   &426.7  &0.17  &478.7 &3721 \\
\hline
$3^{\circ} \times 3^{\circ}$        &2.3   & 675.0  &-3.2  &498.7 &8281 \\
\hline
$4^{\circ} \times 4^{\circ}$        &-0.95   &372.0  &-2.1  &357.6& 14641  \\
\hline
$5^{\circ} \times 5^{\circ}$        &-1.4   & 528.7  &-1.8  &507.7&22801  \\
\hline

\end{tabular}
\end{center}
\end{table}

\begin{figure}
\begin{center}
\vspace{-15pt}
\centerline{\hspace{30pt}\epsfig{file=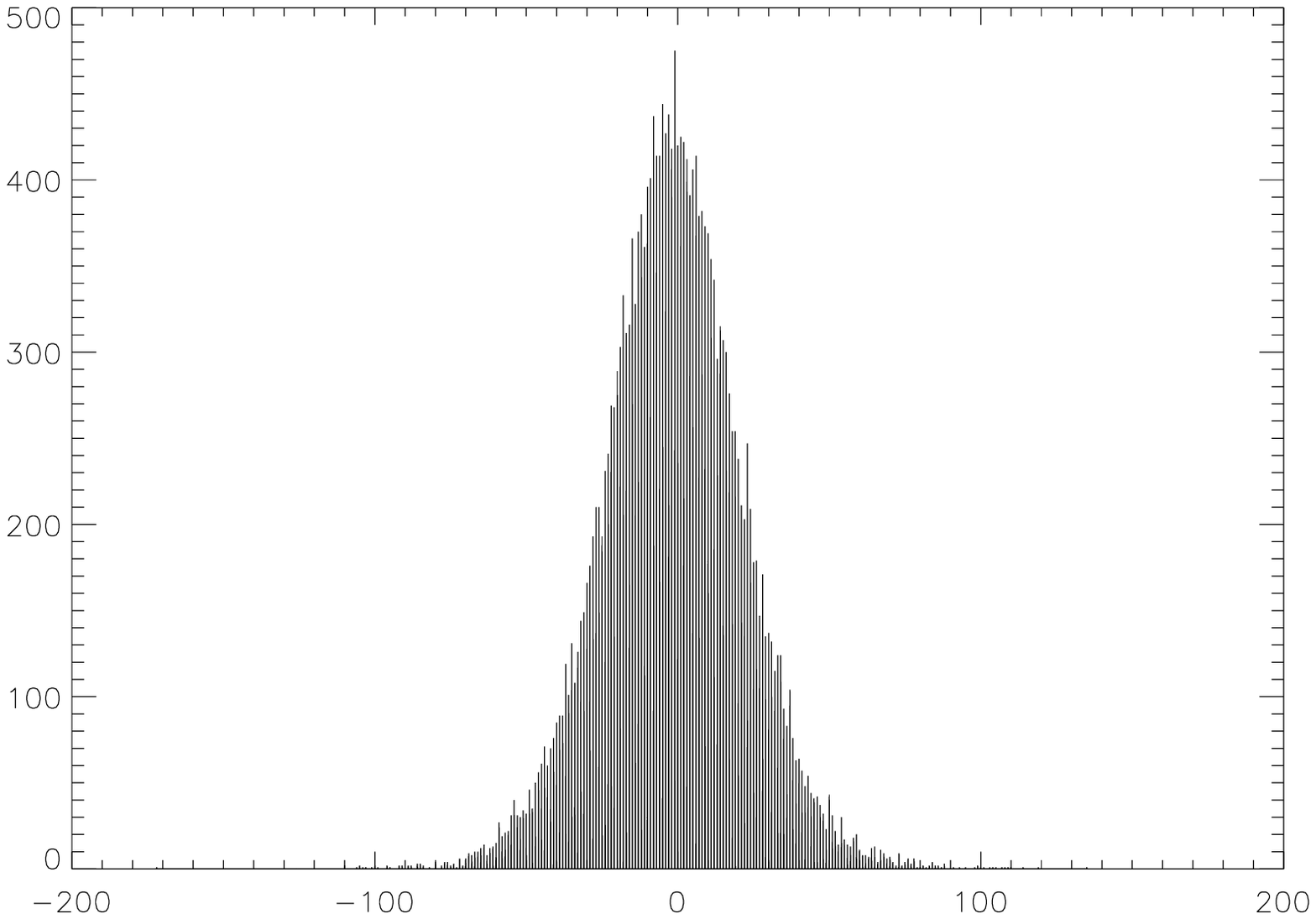,height=4.5in}}
\vspace{-12pt} \caption{\small  Distribution of the values of the Stokes
Parameter $Q$ of the diffuse radiation measured in a $5^{\circ}\times
5^{\circ}$ area of sky within $\pm 5^{\circ}$ from the galactic plane}
\label{fa1}
\end{center}
\end{figure}

\begin{figure}
\begin{center}
\vspace{-15pt}
\centerline{\hspace{30pt}\epsfig{file=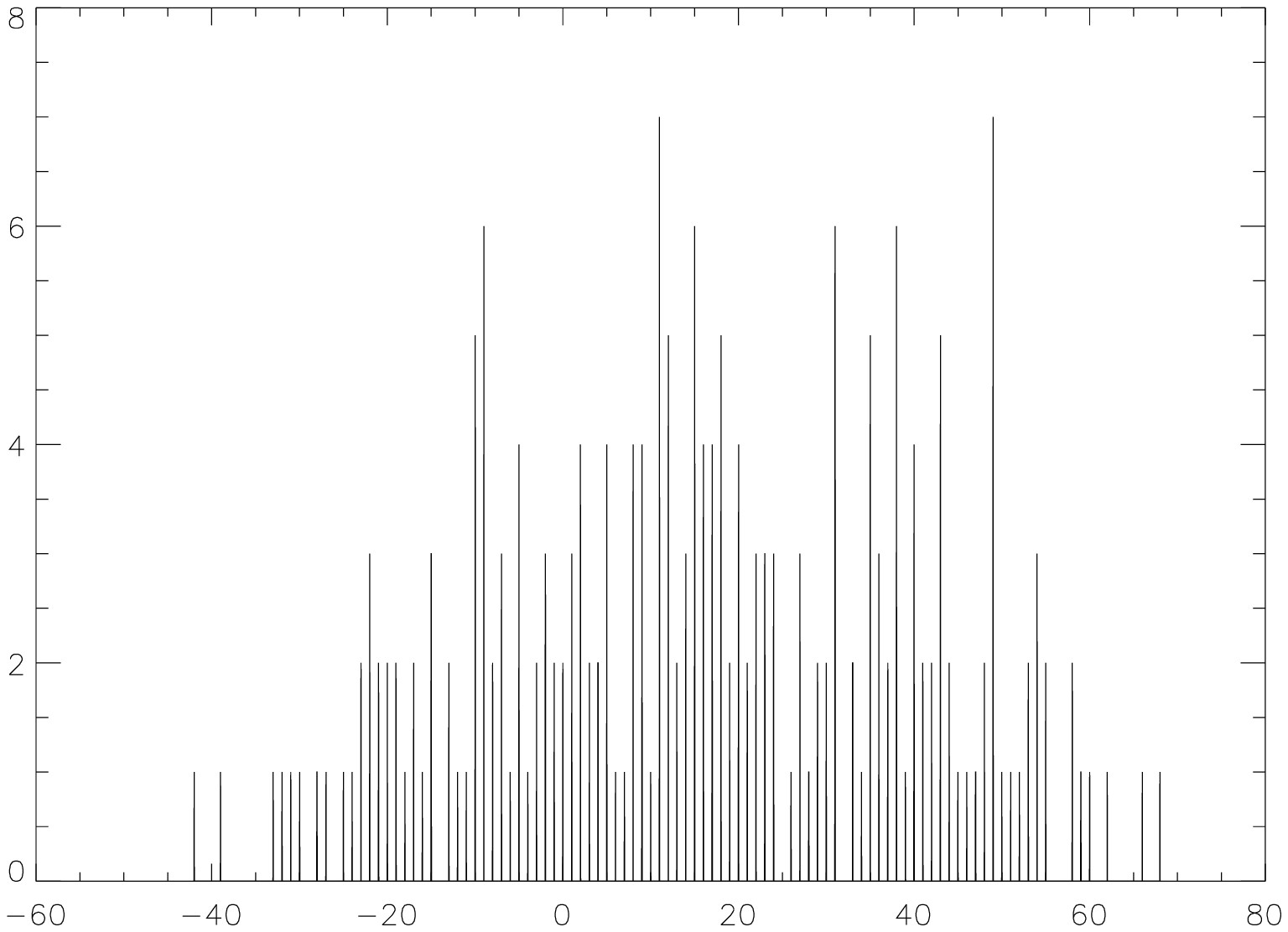,height=4.5in}}
\vspace{-12pt} \caption{\small Distribution of the values of the Stokes
Parameter $Q$ of the diffuse radiation measured in a $0.5^{\circ}\times
0.5^{\circ}$ area of sky within $\pm 5^{\circ}$ from the galactic plane}
\label{fa2}
\end{center}
\end{figure}

\begin{figure}
\begin{center}
\vspace{-15pt} \centerline{\hspace{30pt}\epsfig{file=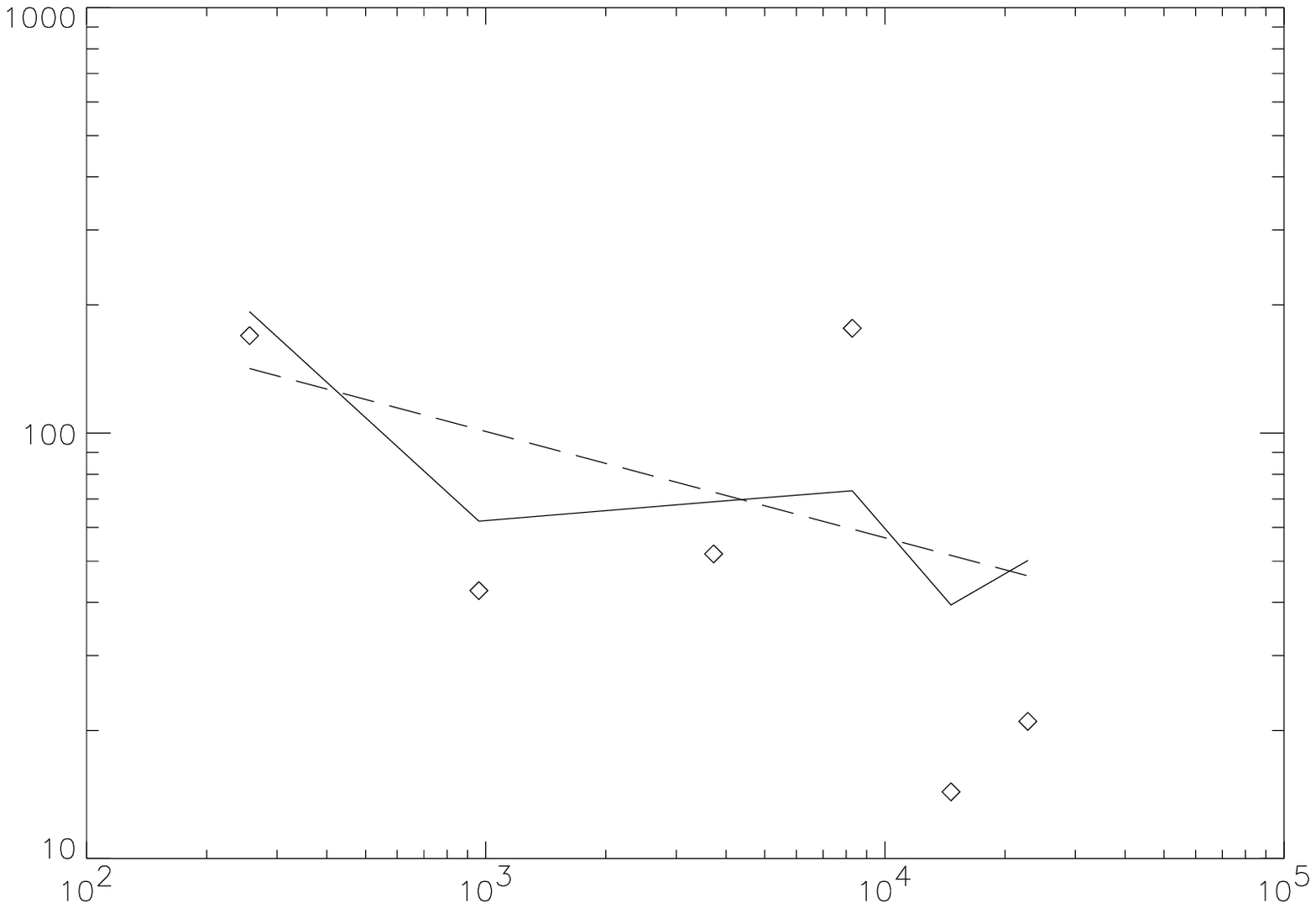,height=4.5in}}
\vspace{-12pt} \caption{\small Plot versus the number $N$ of pixels of the
expected (dashed line, eq.(\ref{appeq1}), $\sigma$ estimated as sum of the
variances of $Q$ and $U$ on all the studied fields) and measured (solid line,
eq. (\ref{appeq2})) values of the variance of our estimator D$D$. Diamonds
are the calculated differences between variances of $Q$ and $U$ and actual
data. (Logarithmic scale are used for clarity). } \label{fa3}
\end{center}
\end{figure}

For each region we examined the shapes of the distributions, and calculated
average value and the variance of the measured values of $Q$ and $U$. The
distribution of $Q$ in a $5^{\circ} \times 5^{\circ}$ area is shown in Figure
\ref{fa1}. Very similar results are obtained for $U$ and for smaller regions.
When the region is very small, (Figure \ref{fa2} is for an $0.5^{\circ} \times
0.5^{\circ}$ area), the distribution becomes broad, but its shape is still
gaussian. Results of analyses carried on on all the areas are presented in
Table 1: for sufficiently large samples both $Q$ and $U$ are compatible with
zero.

To check the validity of eqs. (\ref{var0}) to (\ref{Dl21}) we evaluated the
variance
\begin{equation}
\delta D^2 = \sqrt{2/N} \sigma^2_{Q,U} \label{appeq1}
\end{equation}

\noindent where $N$ is the number of pixels in the field, $\sigma^2_{Q,U}$ is
the measured variance of $Q$ and $U$ and $D\sim \left< Q^2\right> - \left<
U^2\right>$ is our estimator. The quantity  (\ref{appeq1}) is plotted in
Figure \ref{fa3} versus the number $N$ of pixels. In the same figure we plot
also: i)the calculated values of the differences  $ \left< Q^2\right> -
\left< U^2\right>$ in each field and ii)the calculated values of
\begin{equation}
\sqrt{2/N} \frac{\left<(Q - \left<Q\right>)^2\right> - \left<(U -
\left<U\right>)^2\right>}{2} \label{appeq2}
\end{equation}
\par~\par~\par\par


\end{document}